\documentclass[usenatbib,times,usegraphicx,useAMS]{mn2e}
\usepackage{amssymb,url,times,bm}

{\left\lbrace\begin{array}{@{}l@{}}}%
{\end{array}\right.}

\defcitealias{lubow_ogilvie2000}{LO00}
\defcitealias{LOP02}{LOP02}

\title[Wave-like warp propagation in circumbinary discs I]{Wave-like warp propagation in circumbinary discs I.\\ Analytic theory and numerical simulations}

\author[Facchini, Lodato and Price]{Stefano Facchini$^{1,2}$\thanks{facchini@ast.cam.ac.uk}, Giuseppe Lodato$^{1}$ and Daniel J. Price$^3$\\
$^1$Dipartimento di Fisica, Universit\`a Degli Studi di Milano, Via Celoria, 16, Milano, 20133, Italy\\
$^2$Institute of Astronomy, Madingley Road, Cambridge CB3 OHA\\
$^3$Centre for Stellar and Planetary Astrophysics, School of Mathematical Sciences, Monash University, Clayton 3800, Australia.
}
\date{Submission date}
\pagerange{\pageref{firstpage}--\pageref{lastpage}} \pubyear{2012}

\begin{document}
\label{firstpage}
\bibliographystyle{mn2e}
\maketitle

\begin{abstract}

In this paper we analyse the propagation of warps in protostellar circumbinary discs. We use these systems as a test environment in which to study warp propagation in the bending-wave regime, with the addition of an external torque due to the binary gravitational potential. In particular, we want to test the linear regime, for which an analytic theory has been developed. In order to do so, we first compute analytically the steady state shape of an inviscid disc subject to the binary torques. The steady state tilt is a monotonically increasing function of radius, but misalignment is found at the disc inner edge. In the absence of viscosity, the disc does not present any twist. Then, we compare the time-dependent evolution of the warped disc calculated via the known linearised equations both with the analytic solutions and with full 3D numerical simulations. The simulations have been performed with the \textsc{phantom} SPH code using $2$ million particles. We find a good agreement both in the tilt and in the phase evolution for small inclinations, even at very low viscosities. Moreover, we have verified that the linearised equations are able to reproduce the diffusive behaviour when $\alpha>H/R$, where $\alpha$ is the disc viscosity parameter. Finally, we have used the 3D simulations to explore the non-linear regime. We observe a strongly non-linear behaviour, which leads to the breaking of the disc. Then, the inner disc starts precessing with its own precessional frequency. This behaviour has already been observed with numerical simulations in accretion discs around spinning black holes. The evolution of circumstellar accretion discs strongly depends on the warp evolution. Therefore the issue explored in this paper could be of fundamental importance in order to understand the evolution of accretion discs in crowded environments, when the gravitational interaction with other stars is highly likely, and in multiple systems. Moreover, the evolution of the angular momentum of the disc will affect the history of the angular momentum of forming planets.

\end{abstract}



\section{Introduction}
\label{sec:intro}

It is now well known that the majority of stars in star forming regions are in binary or higher order multiple systems \citep[see e.g. the review by][]{mckee_ostriker07}. Moreover, in the last decade the possibility of detecting discs around multiple systems (and both stellar components in close binaries) has dramatically improved. These observations have shown that many stars of this kind do have circumstellar discs and evidence of accretion \citep[e.g.][]{mathieu97}. Therefore, the probability of circumstellar or circumbinary discs around young stars is quite high.

If we focus on the circumbinary case, only a few circumbinary discs have been detected \citep{dutrey94,CM04,beust_dutrey05}. Indirect evidence of their past presence is the recent measure of circumbinary planets \citep{deeg08,lee09,beuermann10}, some of which have been  measured by Kepler \citep[e.g.][]{doyle11,welsh12,orosz12}.

In star forming regions, accretion discs are affected by gravitational interactions with the surroundings \citep{bate10}. These perturbations will strongly affect the evolution of the systems. In particular, the discs are likely to gain a warp, and in the case of multiple systems, to misalign with respect to the stars' orbital plane. This fact has been invoked as a possible explanation of the misalignment between the stellar rotation axis and planets' orbits \citep{bate10}, measurable via the Rossiter McLaughlin effect \citep{triaud10,albrecht12}. In this context, it is of fundamental importance to study the propagation of such interactions into the disc. In particular, warps can be produced by tidal torques due to the binary stars, whenever the binary is misaligned with respect to the disc plane. In this paper we focus on warp propagation in protostellar circumbinary systems in order to take into account these external torques.

Tilted discs have been observed in many other astrophysical environments, such as other galactic binaries as the X-ray binary Her X-1 \citep{tananbaum72,wijers_pringle99}, or the microquasar GRO J1655-40 \citep{hjellming95,martin08}. Warps have also been found in thin discs around AGN, such as NGC 4258 \citep{herrnstein96}. In this last case, additional forcing torques can arise from the general relativistic Lense--Thirring precession around a spinning black hole. The warp evolution is strongly connected with the spin history of the SMBH (Supermassive Black Hole) via the \citet{bardeen_petterson75} effect, and has been studied in depth in recent years \citep[e.g.][]{king08}. Finally, recent studies have been made on spinning SMBH binaries, where the tidal binary torque and the Lense--Thirring effect could coexist \citep{dotti10,nixon2012,lodato_gerosa12}.

Analytic and semi-analytic theories have been developed in order to study warp propagation in two different regimes (see section \ref{sec:waves_theory}): one where the warp evolves diffusively in a thin accretion disc with a diffusive coefficient inversely proportional to disc viscosity \citep{papaloizou_pringle83,pringle92,ogilvie99}, and one where the warp propagates via bending waves in thick or inviscid discs \citep{papaloizou_lin95,lubow_ogilvie2000}.

The diffusive case has been explored in the last years both analytically \citep{scheuer_feiler96,pringle96,LOP02} and numerically \citep{lodato_pringle07,nixon_king12}. In particular, high resolution numerical simulations have been performed, obtaining a good agreement with the analytic theory in the linear and mildly non-linear regime \citep{lodato_price10}.

The bending--wave regime has been less analysed. Analytic studies have given a description of the wave propagation in the linear regime \citep{lubow_ogilvie2000,lubow_ogilvie01,LOP02}, and other studies have been accomplished on the non-linear case \citep{gammie00,ogilvie06}. Only a few poorly resolved numerical simulations have been performed to date via SPH codes \citep{larwood_pap97,nelson_pap99,nelson_pap00}. More recently, \citet{fragner10} performed 3D simulations using a 3D grid code.

In this work we focus on warp propagation via bending waves in protostellar circumbinary discs. After obtaining an analytic solution for the steady state of the warp in a disc extending to infinity, we test the linear regime with numerical simulations. In order to address this issue, we use a 1D ring code \citep[as in][hereafter \citetalias{LOP02}]{LOP02} and full 3D SPH simulations with a much higher resolution than the ones by \citet{nelson_pap99,nelson_pap00}. We find that the agreement between the linear theory and the simulations is good, even at very low viscosities. Finally, 3D simulations allow us to explore the non-linear regime, which has been poorly addressed so far.

While working on this this paper, we found out that a similar problem had been recently studied by \citet{foucart13}. In that paper, they focus on the steady state solutions for circumbinary discs in the linear regime, and analyse the alignment timescale between the disc and the binary. We will compare our respective results when needed.

The paper is organised as follows. In section \ref{sec:potential} we derive a time independent approximation of the gravitational potential generated by the central binary and we extrapolate the torques to which the disc is subjected. In section \ref{sec:waves_theory} we describe warp propagation in the linear regime. In section \ref{sec:analyt} we obtain an analytic solution of the disc steady state in the inviscid limit, and in section \ref{sec:1D} we compare it with 1D time-dependent calculations. In section \ref{sec:res} we describe the numerical setup used to perform the 3D SPH simulations and we report our main results both in  linear and in non-linear regime. Finally, in section \ref{sec:concl} we compare our results with the paper by \citet{foucart13} and draw our conclusions.

In a companion paper \citep{facchini13_2}, we analyse the warp evolution in a specific circumbinary protostellar disc, surrounding the binary system KH 15D \citep{CM04}.

\section{The binary - disc torque}
\label{sec:potential}

In this section we determine the gravitational potential generated by the binary. A similar analysis, restricted to binaries with extreme mass ratios, has been done by \citet{nixon2011}. Here, we generalise their results to arbitrary mass ratios and correct a few typos (compare equation 4 by \citet{nixon2011} with equation \ref{eq:phi_fin}). Note that similar derivations are present in a few other papers, such as \citet{ivanov99} and \citet{naya05}. 

\begin{figure}
\center
\includegraphics[width=.8\columnwidth]{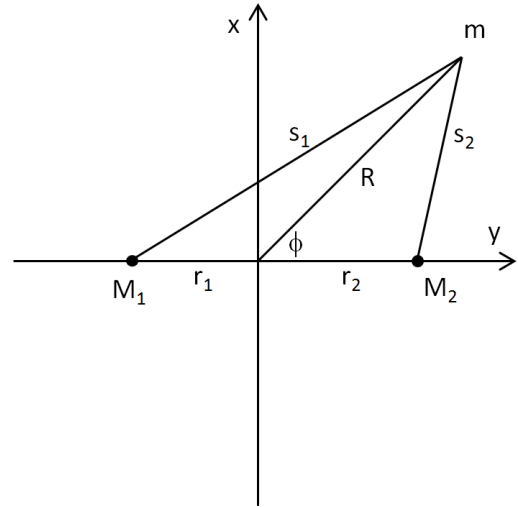}
\caption{The binary system in the corotating frame of reference centred in the centre of mass. $M_1$ and $M_2$ are the two stars, and $m$ is the test particle. For simplicity in this figure we locate $m$ in the $(R,\phi)$ plane. The quantity $r_1$ indicates the distance between $M_1$ and the centre of mass, and analogously for $r_2$ and $M_2$. The distances between $m-M_1$ and $m-M_2$ are called $s_1$ and $s_2$, respectively.}
\label{fig:grafico_pot}
\end{figure}

We consider two stars with masses $M_1$ and $M_2$, that rotate in two circular orbits, on a plane described by the cylindrical polar coordinates $(R,\phi)$ and perpendicular to the $z-$axis. We place the origin of our coordinate system in the centre of mass. The geometry of the system, as well as the definition of various quantities of interest, is shown in Fig. \ref{fig:grafico_pot}. We denote with $r_1$ the distance between $M_1$ and the centre of mass, and similarly for $r_2$ and $M_2$. The distance between the two stars is $a=r_1+r_2$. We finally place a test particle $m$ in a generic position with coordinates $(R,\phi,z)$. We call $s_1$ and $s_2$ the distance between $m$ and $M_1$ and $m$ and $M_2$.  We are focusing on a restricted three body problem in 3D space.

We now consider a reference frame $S$ corotating with the two stars. In this particular system they are both at rest at $(r_1,\pi,0)$ and $(r_2,0,0)$, respectively. The potential in this reference frame is:
\begin{equation}
\label{eq:phi_init}
\Phi(R,\phi,z)=-G\left(\frac{M_1}{s_1} + \frac{M_2}{s_2} \right) - \frac{1}{2}\Omega_{\rm b}^2R^2,
\end{equation}
where $\Omega_{\rm b}$ is the angular velocity of the binary, given by
\begin{equation}
\Omega_{\rm b}^2=\frac{G(M_1+M_2)}{a^3}.
\end{equation}
In equation \ref{eq:phi_init} the second term is due to the non-inertial nature of the reference frame, and it represents the centrifugal potential. The orbital frequency $\Omega_{\rm b}$ is simply calculated from Kepler's third law. Finally it can be easily shown that:
\begin{eqnarray}
\nonumber s_1^2 = r_1^2 + R^2 + 2r_1R\cos{\phi} + z^2;\\
s_2^2 = r_2^2 + R^2 - 2r_2R\cos{\phi} + z^2.
\end{eqnarray}
We now move to an inertial reference frame $S'$, with the origin coincident with that of $S$.  We do not have the centrifugal term anymore, and the $\phi$ angle undergoes the simple transformation $\phi\rightarrow\phi '=\Omega_{\rm b} t$. In this frame we obtain the following gravitational potential:
\begin{eqnarray}
\nonumber\Phi(R,\phi'=\Omega_{\rm b} t,z) =& -G \displaystyle \frac{M_1}{(R^2 + r_1^2 + 2r_1R\cos{\Omega_bt} + z^2)^{1/2}}\\
& - G \displaystyle \frac{M_2}{(R^2 + r_2^2 - 2r_2R\cos{\Omega_b t} + z^2)^{1/2}}.
\end{eqnarray}
This is the most general form of the gravitational potential of a circular binary. We could now expand the above relation in a Fourier series with azimuthal wavenumber $\sigma$ to distinguish the various contributions to the potential. In order to avoid it, we make the assumption, as it has been made by \citet{nixon2011} and earlier by e.g. \citet{lubow_ogilvie01}, that the perturbations to the potential of the $\sigma\geq1$ modes are oscillatory, and if we are far enough from resonances, they will have no long-term secular effect. Long-term effects on the orbit of the test particle, and hence eventually on the disc, come from the only zero-frequency term. Since we are interested in the secular dynamics of the disc, we can just consider this $\sigma=0$ mode \citep{bate00}.

\begin{figure}
\center
\includegraphics[width=.8\columnwidth]{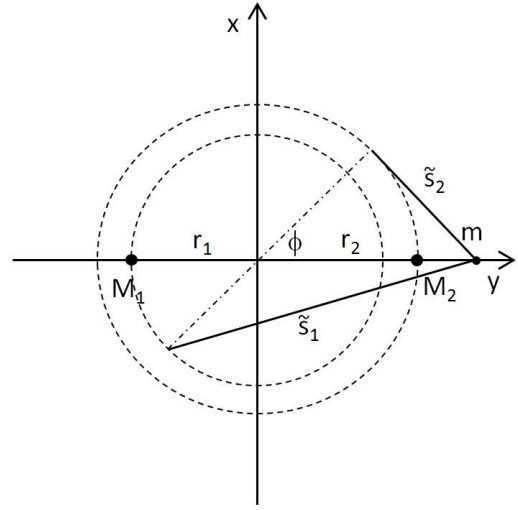}
\caption{The binary system in an inertial frame of reference centred in the centre of mass. $M_1$ and $M_2$ are now spread in two massive annuli with radii $r_1$ and $r_2$, respectively, and $m$ is the test particle, positioned at $(R,0,z)$ (we place it at $\phi=0$ because of the rotational symmetry about the $z$-axis of the problem). The new $\tilde{s}_1$ and $\tilde{s}_2$ define the distances between a generic point of the two massive annuli and the test particle.}
\label{fig:grafico_pot_2}
\end{figure}

In order to calculate this time-independent term we use the fact that physically this $\sigma=0$ term is given by replacing the two masses $M_1$ and $M_2$ with the same masses spread uniformly over their orbit, i.e. two rings of mass $M_1$ and $M_2$ and radius $r_1$ and $r_2$ in the $(R,\phi)$ plane \citep[see][]{nixon2011}. The gravitational potential for the generic test particle is then:
\begin{equation}
\label{eq:phi_m1_m2}
\Phi(R,z)=-\frac{GM_1}{2\pi}\int_0^{2\pi}{\frac{d\phi}{\tilde{s}_1}}-\frac{GM_2}{2\pi}\int_0^{2\pi}{\frac{d\phi}{\tilde{s}_2}},
\end{equation}
where
\begin{eqnarray}
\nonumber\tilde{s}_1^2=R^2+r_1^2+z^2+2Rr_1\cos{\phi}, \\
\tilde{s}_2^2=R^2+r_2^2+z^2-2Rr_2\cos{\phi}.
\end{eqnarray}
The new $\tilde{s}_1$ and $\tilde{s}_2$ define the distances between a generic point of the two massive annuli and the test particle positioned at $(R,0,z)$ (we place it at $\phi=0$ because of the rotational symmetry about the $z$-axis of the problem, see Fig. \ref{fig:grafico_pot_2}), while $s_1$ and $s_2$ indicate the distances between the two stars at rest in the corotating reference frame and a generic particle.

We introduce the factor $\eta=M_1M_2/M^2$ and the total mass $M=M_1+M_2$. We recall that $a=r_1+r_2$. If we now expand equation \ref{eq:phi_m1_m2} in powers of $r_1/R$, $r_2/R$ and $z/R$, keeping terms only up to second order we find:
\begin{equation}
\label{eq:phi_fin}
\nonumber\Phi(R,z)=-\frac{GM}{R}-\frac{GM\eta a^2}{4R^3} + \frac{GMz^2}{2R^3}+\frac{9}{8}\frac{GM\eta a^2z^2}{R^5}.
\end{equation}
We notice that by taking the limits $\eta\rightarrow M_2/M_1$ and $M\rightarrow M_1$ we obtain the equivalent of equation $4$ of \citet{nixon2011}, in which they calculated the same gravitational potential but in the simplified case where $M_2\ll M_1$.

The perturbations affecting the disc particles can be expressed in terms of the orbital frequency $\Omega$, the vertical oscillation frequency $\Omega_z$, and the epicyclic frequency $\kappa$ (see section \ref{sec:waves_theory}). Their definition follows below:
\begin{equation}
\Omega^2=\frac{1}{R}\frac{\partial\Phi}{\partial R}\bigg|_{z=0},
\end{equation}
\begin{equation}
\Omega_z^2=\frac{\partial^2\Phi}{\partial z^2}\bigg|_{z=0},
\end{equation}
\begin{equation}
\kappa^2=4\Omega^2+2R\Omega\frac{d \Omega}{d R}=4\Omega^2\left[1+\frac{1}{2}\frac{d\ln{\Omega}}{d\ln{R}}\right].
\end{equation}

For the binary potential of equation \ref{eq:phi_fin} we obtain:
\begin{equation}
\label{eq:omega}
\Omega^2=\frac{GM}{R^3}+\frac{3}{4}\frac{GM\eta a^2}{R^5},
\end{equation}
\begin{equation}
\Omega_z^2=\frac{GM}{R^3}+\frac{9}{4}\frac{GM\eta a^2}{R^5},
\end{equation}
\begin{equation}
\label{eq:kappa}
\kappa^2=\frac{GM}{R^3}-\frac{3}{4}\frac{GM\eta a^2}{R^5}.
\end{equation}
To first order, thus,
\begin{equation}
\label{eq:omegaz2}
\frac{\Omega_z^2-\Omega^2}{\Omega^2}=\frac{3}{2}\frac{\eta a^2}{R^2},
\end{equation}
\begin{equation}
\label{eq:kappa2}
\frac{\kappa^2-\Omega^2}{\Omega^2}=-\frac{3}{2}\frac{\eta a^2}{R^2}.
\end{equation}
We will use this approximation throughout the paper.

In summary, in this section we have deduced the time independent term of the gravitational potential of a generic binary system, formed by two different stars of mass $M_1$ and $M_2$, in the hypothesis that the disc particles rotate at a large radius ($R\gg r_1,\ r_2$) and low height ($z\ll R$).

\section{Theory of wave-like warp propagation}
\label{sec:waves_theory}

We consider here the propagation of warps in thin, almost Keplerian accretion discs. The quantities that define their dynamics are the angular velocity $\Omega(R)$, the surface density $\Sigma(R)$ and the angular momentum per unit area ${\bf L}(R)$.  $H$ describes the scale height of the disc, and it is related to the sound speed $c_{\rm s}$ via $H=c_{\rm s}/\Omega$. We assume here that the disc is composed of a series of flat, infinitesimally thin rings, each of which can be oriented arbitrarily in space. A single ring at radius $R$ is thus described by two angles: the tilt angle $\beta$ with respect to the $z$ axis, and the azimuthal angle $\gamma$ that defines the orientation of the tilt with respect to an arbitrary axis, perpendicular to $z$. If $\beta$ varies with $R$, we will have a warped disc. If $\gamma$ varies with radius the disc is additionally twisted. Here, $R$ should be intended as a spherical coordinate, even though at each radius the disc is thin in the direction perpendicular to the local rotation plane. Therefore we define in complex notation the tilt of the disc $W(R,t)$ at each radius as $W(R,t)=\beta(R,t)\exp{[i\gamma(R,t)]}$ \citep{pringle96}. These quantities are related to the specific angular momentum through ${\bf l}(R) = {\bf L}(R)/L(R) = (\cos\gamma\sin\beta,\sin\gamma\sin\beta,\cos\beta)$. 

We consider a standard $\alpha$-prescription for the viscosity: $\nu=\alpha c_{\rm s}H$ \citep{shakura73}.

Warp propagation can be described in two different regimes. \citet{papaloizou_pringle83} had already suggested that whenever $\alpha < H/R < 1$ warps would probably propagate via bending waves, whereas when $H/R < \alpha < 1$ the equations describing the evolution would be diffusive \citep{pringle92}. These results have been confirmed analytically; \citet{papaloizou_lin95} derived the equations describing the evolution in the case $\alpha < H/R < 1$, and they confirmed that they evolve via wave equations. Equivalent formulations have been derived later by \citet{demianski_ivanov97} and \citet{lubow_ogilvie2000} (hereafter, \citetalias{lubow_ogilvie2000}). Throughout this work we use the formulation by \citetalias{lubow_ogilvie2000}.

\citetalias{lubow_ogilvie2000} have shown that when the disc is nearly Keplerian and non self-gravitating, the linearised equations for bending waves (with azimuthal wavenumber $m=1$) may be written as:
\begin{equation}
\label{eq:wave_l_real}
\Sigma R^2\Omega\frac{\partial {\bf l}}{\partial t}=\frac{1}{R}\frac{\partial {\bf G}}{\partial R}+{\bf T},
\end{equation}
and
\begin{equation}
\label{eq:wave_g_real}
\frac{\partial{\bf G}}{\partial t}+\left(\frac{\kappa^2-\Omega^2}{\Omega^2}\right)\frac{\Omega}{2}{\bf e}_z\times{\bf G} +\alpha\Omega{\bf G}=\Sigma R^3\Omega\frac{c_{\mathrm{s}}^2}{4}\frac{\partial{\bf l}}{\partial R},
\end{equation}
where
\begin{equation}
\label{eq:external_torque}
{\bf T}=-\Sigma R^2\Omega\left(\frac{\Omega_z^2-\Omega^2}{\Omega^2}\right)\frac{\Omega}{2}{\bf e}_z\times{\bf l},
\end{equation}
where ${\bf e}_z$ is the unit vector perpendicular to the binary orbit.

In our assumptions the warp is small, therefore $l_x,l_y \ll 1$ and $l_z \approx 1$. Thus, by considering $l_z = 1$, we can consider the equations on the $xy$-plane only. The term $2\pi {\bf G}$ is the internal torque and ${\bf T}$ the external torque density. The external torque is due to the lack of spherical symmetry in the potential, and is proportional to the term $\Omega_z^2 - \Omega^2$ (equation \ref{eq:external_torque}). Instead, equation \ref{eq:wave_g_real} shows that the internal torque is mediated by horizontal epicyclic motions. The term proportional to $\alpha$ tends to dissipate the waves through an exponential factor. Finally, note that the external torque is related to the precession frequency of the ring ${\bf \Omega}_{\rm p}$, as we know it should be from simple mechanics \citep{lodato_pringle06}. By knowing that ${\bf L}=\Sigma R^2 \Omega {\bf l}$ we can rewrite equation \ref{eq:external_torque} as:
\begin{equation}
{\bf T}=\frac{-(\Omega_z-\Omega)(\Omega_z+\Omega)}{\Omega^2}\frac{\Omega}{2}{\bf e}_z \times {\bf L} \approx {\bf \Omega}_{\rm p}\times{\bf L}.
\end{equation}
If $\Omega_z\approx\Omega$, then ${\bf \Omega}_{\rm p}=(\Omega-\Omega_z){\bf e}_z$ \citep{nixon2011}.

A different but equivalent set of equations can be used by defining the dimensionless complex variable $W(R,t)=l_x+il_y$ and the complex variable $G(R,t)=G_x+iG_y$. We can thus rewrite equation \ref{eq:wave_l_real} and \ref{eq:wave_g_real} as:

\begin{equation}
\label{eq:wave_l_complex}
\Sigma R^2\Omega\left[\frac{\partial W}{\partial t}+\left(\frac{\Omega_z^2-\Omega^2}{\Omega^2}\right)\frac{i\Omega}{2}W \right]=\frac{1}{R}\frac{\partial G}{\partial R},
\end{equation}
and

\begin{equation}
\label{eq:wave_g_complex}
\frac{\partial G}{\partial t}+\left(\frac{\kappa^2-\Omega^2}{\Omega^2}\right)\frac{i\Omega}{2}G+\alpha\Omega G=\Sigma R^3\Omega\frac{c_{\mathrm{s}}^2}{4}\frac{\partial W}{\partial R}.
\end{equation}

Let us consider the propagation velocity of the waves. By neglecting the external torque and the non-Keplerian term we can obtain a first order approximated dispersion relation \citep{nelson_pap99}:

\begin{equation}
\omega=\frac{1}{2}[i\alpha\Omega \pm (c_{\rm s}^2k^2 - \alpha^2\Omega^2)^{\frac{1}{2}}],
\end{equation}
where $k$ is the radial wavenumber and $\omega$ is the wave frequency. Therefore if the disc is inviscid ($\alpha=0$) the warp propagates as a non-dispersive wave with wave speed $c_{\rm s}/2$. Moreover, propagation becomes purely diffusive in the limit $|\omega| \ll \alpha\Omega$. Note that in the Keplerian limit ($\Omega=\Omega_z=\kappa$) equations \ref{eq:wave_l_real} and \ref{eq:wave_g_real} explicitly tend to a diffusive equation when this condition is verified (i.e. when the third term of the l.h.s. of equation \ref{eq:wave_g_real} dominates over the first two).

Finally, \citetalias{LOP02} have shown that in the inviscid case the dispersion relation associated to equation \ref{eq:wave_l_real} and \ref{eq:wave_g_real} is given by:
\begin{equation}
\left[ \omega - \left( \frac{\Omega^2-\Omega_z^2}{2\Omega}\right)\right] \left[ \omega - \left( \frac{\Omega^2-\kappa^2}{2\Omega}\right)\right]=\frac{c_{\rm s}^2}{4}k^2.
\end{equation}
In the case of $\omega=0$, whenever $\kappa^2-2\Omega^2+\Omega_z^2=0$, which is the case for our binary potential, the spatial configuration that the disc will reach is an evanescent wave. We shall see that this theoretical prediction made by \citetalias{LOP02} is verified by our results.

Finally, note that in this section we have considered the linear case only. Some efforts have been spent in the last decade to cover the non-linear case in Keplerian and nearly Keplerian discs, both in the diffusive regime \citep{ogilvie99,lodato_price10} and in the wave-like one \citep{ogilvie06}.

\section{Analytic considerations for the steady state}
\label{sec:analyt}

In this section we deduce an analytic solution for the warped disc shape in its steady state. We follow the procedure by \citetalias{LOP02}, who considered the solution for an external torque due to a spinning black hole. We consider a circumbinary disc extending from $R_{\rm in}$ up to $R_{\rm out}\rightarrow\infty$, under all the approximations made in section \ref{sec:potential}. Therefore we can implement equations (\ref{eq:omega}-\ref{eq:kappa}) in equations (\ref{eq:wave_l_complex}) and \ref{eq:wave_g_complex}. By setting the time-derivatives to $0$, we obtain equation $17$ of \citetalias{LOP02}:
\begin{equation}
\frac{d}{dR}\left[\left(\frac{\Sigma c_{\mathrm{s}}^2R^3\Omega^2}{\Omega^2-\kappa^2+2i\alpha\Omega}\right)\frac{dW}{dR}\right] +\Sigma R^3(\Omega^2-\Omega_z^2)W=0.
\end{equation}
We focus on the inviscid case. In order to explicitly write the above equation, we consider $\Sigma\propto R^{-p}$ and $c_{\rm s}\propto R^{-q}$. We use the parametrisation $R=R_{\rm in}x$. We obtain
\begin{equation}
\label{eq:analyt}
\frac{d}{dx}\left[x^{5-2q-p}\frac{dW}{dx}\right]=4\chi^2x^{-2-p}W,
\end{equation}
where
\begin{equation}
\chi=\frac{3}{4}\eta\frac{(a/R_{\rm in})^2}{H_{\rm in}/R_{\rm in}}.
\label{eq:defchi}
\end{equation}
Note that equation \ref{eq:analyt} can be projected to the real domain only. The real and the imaginary parts of $W$ are not entangled anymore. This has the important consequence that the steady state shape of an inviscid disc will have no twist.

In the absence of viscosity we have just two phenomena that contrast each other: gravity and pressure. The parameter $\chi$ indicates which one dominates. Note that the amplitude of the warp will be regulated by the $\chi$ parameter only (cf. equation 20 of \citet{foucart13}).

After some algebra, we can express the solutions of equation \ref{eq:analyt} in terms of modified Bessel functions of the first and second kind:

\begin{equation}
W(x)=x^{-\zeta}[c_1 I_{\xi}(y(x))+c_2 K_{\xi}(y(x))],
\end{equation}
where
\begin{equation}
y(x)=\left|\frac{2\chi x^{\psi}}{\psi}\right|,
\end{equation}
and
\begin{equation}
\zeta=-\frac{1}{2}(2q+p-4),\ \ \ \ \psi=\frac{1}{2}(2q-5),\ \ \ \ \xi = \frac{\zeta}{\psi}.
\end{equation}
This is the most general solution for the steady shape of an inviscid disc around a circular binary system. The modified Bessel functions $I$ and $K$ are exponentially growing and decaying functions and are not oscillatory. This confirms the theoretical prediction highlighted by \citetalias{LOP02} and reported at the end of section \ref{sec:waves_theory} in the case of $\omega=0$. This is a relevant difference between the binary case and the steady state of a disc under Lense-Thirring torques, where the solutions are oscillatory.

Finally, we need to specify the boundary conditions. We set $\lim_{x\rightarrow\infty}{W(x)}=W_{\infty}$, which comes from the fact that $W$ has an horizontal asymptote (at large radii the disc is not affected by the external torque, which decreases as $R^{-7/2}$) and $W'(x=1)=0$ (zero-torque boundary condition at the inner edge). In this way we constrain the two constants $c_1$ and $c_2$ \footnote{Note that in order to have a finite tilt angle for $R\rightarrow\infty$, we need to require that $\zeta>0$. This is generally satisfied, unless either the sound speed or the surface density is a very steeply decaying function of $R$. In these cases, the warp cannot be communicated effectively through the disc, which will behave essentially as a single precessing ring, and will thus not reach a steady (non-precessing) shape.}. Note that $W_{\infty}$ is an arbitrary number, that in real cases will depend on the initial condition of the disc tilt. All solutions are valid to within a constant scale amplitude, since we deal with a linear model. To compute the limits, we use the fact that when $r\rightarrow 0$:
\begin{equation}
I_{\xi}(y)\sim\frac{1}{\Gamma(\xi+1)}\left(\frac{y}{2}\right)^{\xi},
\end{equation}
and, in general: 
\begin{equation}
K_{\xi}(y)=\frac{\pi[I_{-\xi}(y)-I_{\xi}(y)]}{2\sin{\xi\pi}},
\end{equation}
where $\Gamma$ is the gamma function. We do not need any other condition, in particular we do not need to fix a value for $W$ at the inner edge of the disc. In general the tilt at the inner edge of the disc will not be very small. This is an importance difference between the wave-like and the diffusive regime, which tends to have a very small tilt in the inner regions of the disc.

\begin{figure}
\begin{center}
\includegraphics[width=.85\columnwidth]{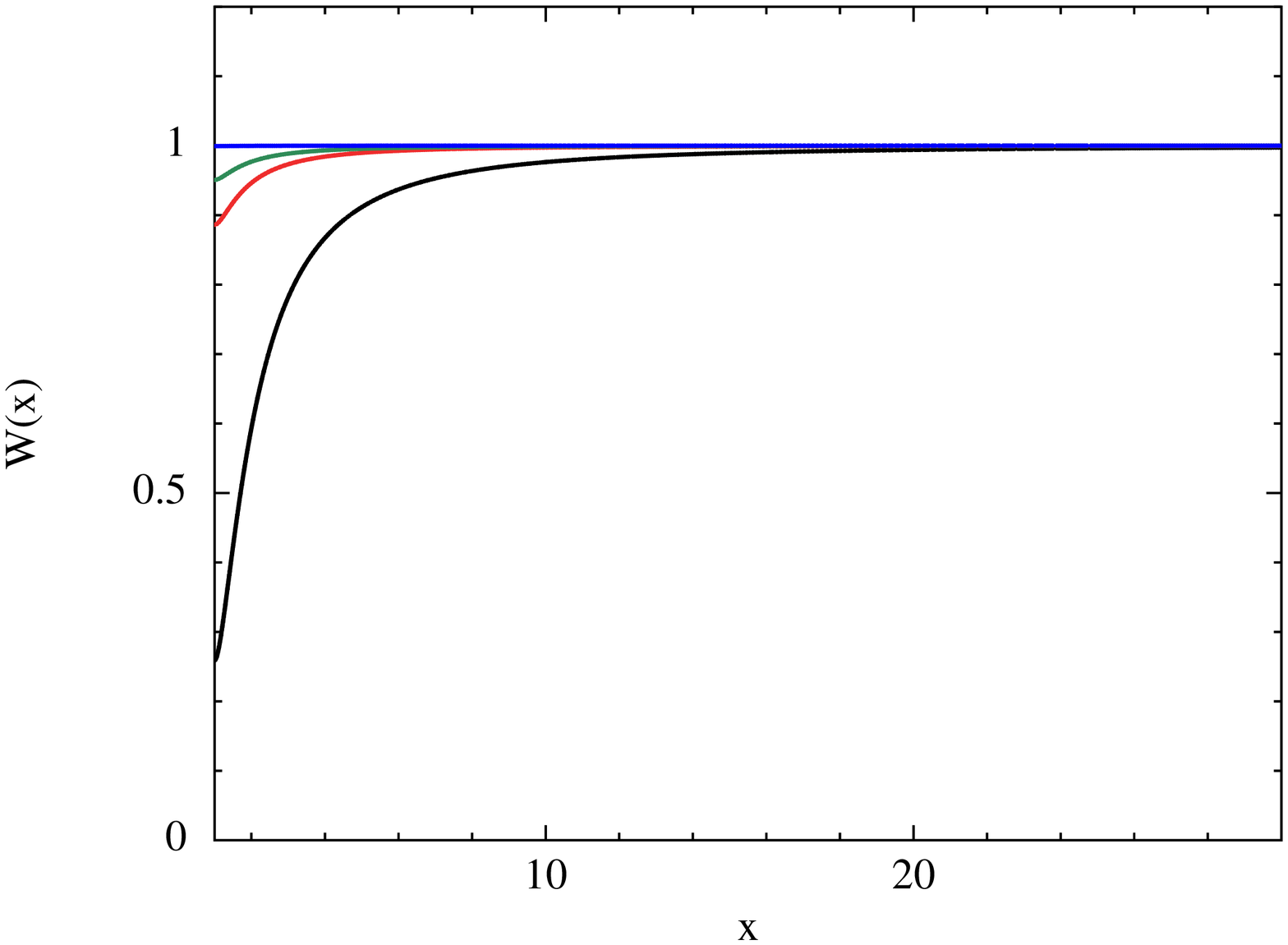}
\includegraphics[width=.85\columnwidth]{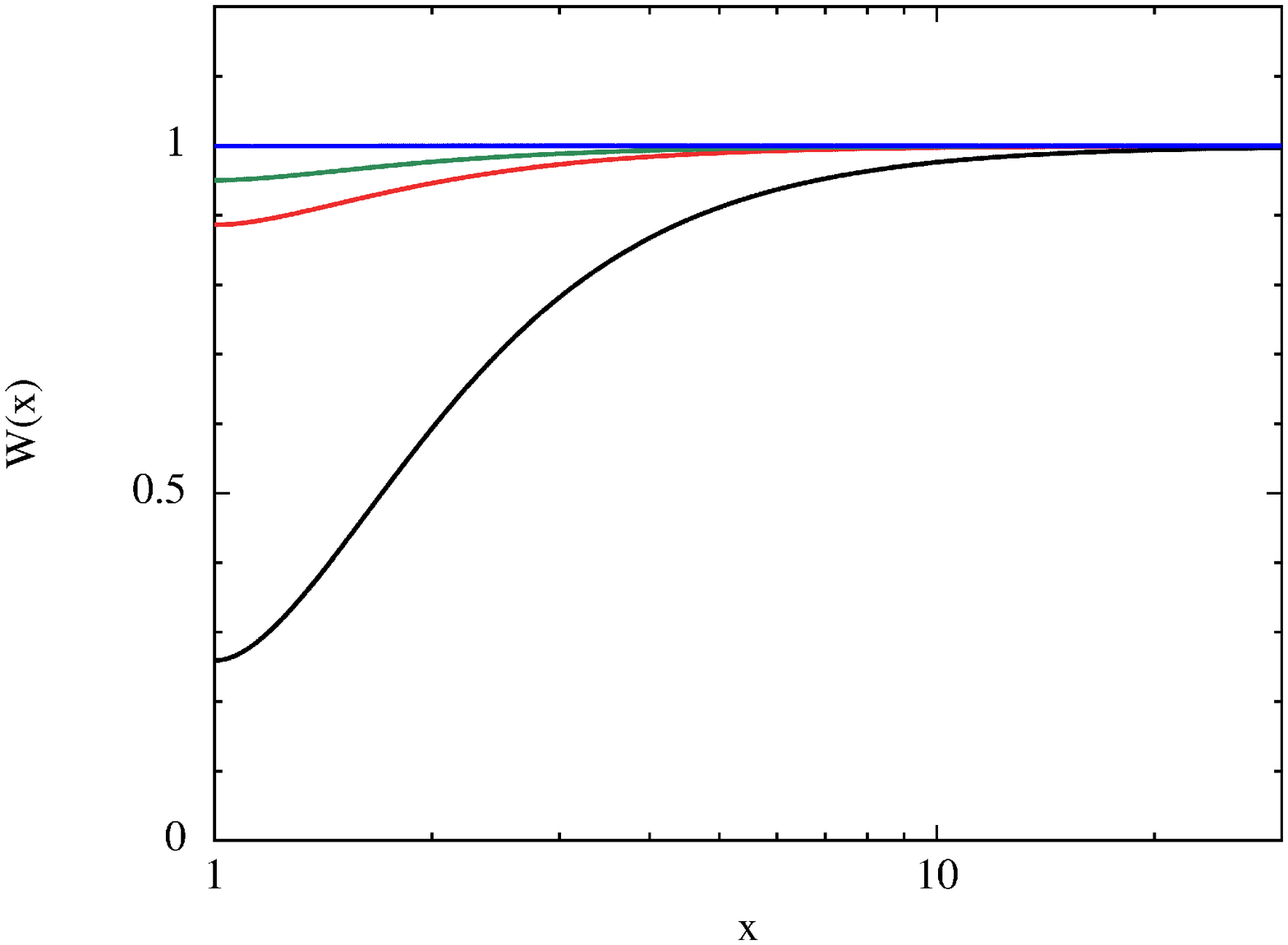}
\end{center}
\caption{Analytic solution of $W(x)$ with $H_{\mathrm{in}}/R_{\mathrm{in}}=0.1$, $p=0.5$ and $q=0.75$. The bottom panel shows the same plot as in the top one with a logarithmically scaled $x-$axis. The three coloured lines (blue, green, red) illustrate the case in which $R_{\rm in}=2a$, and $\eta=0.0475$, $0.16$ and $0.25$, respectively. Note that the amplitude of the warp increases with the binary mass ratio $M_2/M_1<1$. The black line shows the case $\eta=0.25$ and $R_{\rm in}=a$. In all the solutions $W_{\infty}$ has been set to $1$. The general solution is an evanescent wave, without any oscillation.}
\label{fig:analytical}
\end{figure}

So far we have used two generic values for $q$ and $p$. Henceforth we set the two values to $3/4$ and $1/2$, respectively. In Fig. \ref{fig:analytical} we report the solution obtained with a few sets of typical physical parameters for protostellar circumbinary discs. The scale height is equal in all the portrayed solutions: $H_{\mathrm{in}}/R_{\mathrm{in}}=0.1$, and scales as $H/R\propto R^{-1/4}$. From \citet{art_lubow94} we know that the inner radius of the disc is equal to the tidal truncation radius, that in the case of circular orbits is $R_{\rm t}\sim 2a$ (the dependence on $\eta$ is very weak). In Fig. \ref{fig:analytical} we set $R_{\rm in}=2a$ for the three coloured (blue, green, red) lines, which illustrate the analytic solution for the following values of $\eta$: $0.0475$, $0.16$ and $0.25$ (corresponding to the mass ratio $M_2/M_1 = 0.052, 0.25$ and 1, respectively).The black line shows the case in which $\eta=0.25$, but $R_{\rm in}=a$ so that the disc extends to an inner radius that is slightly smaller than the tidal truncation radius. Note that, for the same inner radius, the warp becomes more prominent as the mass ratio between the two stars becomes closer to 1 ($\eta\rightarrow 0.25$), but still maintains a significant misalignment even for equal masses. For the same mass ratio, obviously, the warp increases as the disc moves closer in towards the binary orbit. We show the $R_{\rm in}=a$ case to underline the strong dependence of the warping on the inner radius of the disc. Finally, we have verified that the order of magnitude of the amplitude of the warp agrees with the estimate given by equation 20 by \citet{foucart13} \citep[compare the case $\eta=0.25$, $R_{\rm in}=2a$, i.e. the red line in Fig. \ref{fig:analytical}, with the tilting showed in fig. 1 by][]{foucart13}.

\section{Time-dependent evolution: a 1D model}
\label{sec:1D}
In this section we describe and use a 1D model for warp propagation via bending waves in a disc subject to a binary torque. We consider $R$ as the only spatial variable of the system, as described at the beginning of section \ref{sec:analyt}. The disc is discretised into a set of thin annuli that can be tilted and interact with one another via pressure and viscous forces. In this dynamical evolution, we neglect the dependence of $\Sigma$ on time. In fact, from the dispersion relation of the wave equations we know that for low viscosity discs, bending waves propagate on a timescale $t_{\rm dyn}=2R/c_{\rm s}$, whereas the viscous evolution of $\Sigma$ occurs on a timescale $t_{\nu}=R^2/\nu$. Therefore:
\begin{equation}
\frac{t_{\rm dyn}}{t_{\nu}}=\frac{2R\nu}{c_{\rm s}R^2}=2\alpha\frac{H}{R}.
\end{equation}
Since we know that $\alpha<H/R\ll 1$ we can neglect the evolution of $\Sigma$.

Moreover, we neglect the angular momentum variations of the binary, which is affected by the gravitational potential of the disc. We do not consider the back-reaction of the disc onto the binary angular momentum because we focus on low mass discs, where their angular momentum is negligible compared to the binary one (this back-reaction is considered in \citealt{foucart13}). 

In order to compute the evolution we move to four dimensionless differential equations from equations \ref{eq:wave_l_real} and \ref{eq:wave_g_real}. We use the following parametrisation: $R=R_{\rm in}x$, $\Omega=\Omega_{\rm in}x^{-3/2}$, $\Sigma=\Sigma_{\rm in}x^{-p}$, $t=\Omega_{\rm in}^{-1}(H_{\rm in}/R_{\rm in})^{-1}\tau$, ${\bf l}=W_{\infty}\boldsymbol{\lambda}$ and ${\bf G}=G_{\rm in}{\bf \Gamma}$. Then, we set $G_{\rm in}=\Sigma_{\rm in}R_{\rm in}^4\Omega_{\rm in}^2(H_{\rm in}/R_{\rm in})W_{\infty}$. With these definitions, we obtain the following set of 4 equations:
\begin{equation}
\label{eq:lambda_1}
\frac{\partial\lambda_x}{\partial \tau}=x^{p-3/2}\frac{\partial\Gamma_x}{\partial x}+\chi x^{-7/2}\lambda_y,
\end{equation}
\begin{equation}
\frac{\partial\lambda_y}{\partial \tau}=x^{p-3/2}\frac{\partial\Gamma_y}{\partial x}-\chi x^{-7/2}\lambda_y,
\end{equation}
\begin{equation}
\frac{\partial\Gamma_x}{\partial \tau}+\alpha \frac{R_{\mathrm{in}}}{H_{\mathrm{in}}} x^{-3/2}\Gamma_x +\chi x^{-7/2}\Gamma_y=x^{3/2-p}\left(\frac{c}{2}\right)^2\frac{\partial\lambda_x}{\partial x},
\end{equation}
\begin{equation}
\label{eq:gamma_2}
\frac{\partial\Gamma_y}{\partial \tau}+\alpha \frac{R_{\mathrm{in}}}{H_{\mathrm{in}}} x^{-3/2}\Gamma_y -\chi x^{-7/2}\Gamma_x=x^{3/2-p}\left(\frac{c}{2}\right)^2\frac{\partial\lambda_y}{\partial x},
\end{equation}
where $c$ is the dimensionless sound speed ($c=x^{-3/4}$ in this paper), and $\chi$ has been previously defined in equation \ref{eq:defchi}. All the physics is included in two parameters: $\alpha/(H_{\rm in}/R_{\rm in})$ and $\chi$. The first one is the measure of the importance of viscous and pressure effects, the second one determines the magnitude of the external torque due to the binary potential with respect to pressure forces. The parameter $H_{\rm in}/R_{\rm in}$ is a scale parameter determining the speed of the temporal evolution.

\citetalias{LOP02} solved the same problem for the Lense-Thirring case with a different but equivalent set of equations in the complex domain. We prefer to use our equations, because they show the dependence of the evolution on the physical parameters more transparently. However, we use their result as an important comparison. By implementing the Lense-Thirring torque, and by setting the exact same set of parameters as they did, we obtain their same result for the disc tilt shape. This confirms the equivalence of the two sets of equations.

\subsection{Code and boundary conditions}

The numerical code we use to solve equations \ref{eq:lambda_1}-\ref{eq:gamma_2} implements the same numerical algorithm used by \citetalias{LOP02}. We consider ${\bf \Gamma}$ to be defined at $N$ logarithmically distributed grid points (typically $N=1001$, for the inviscid simulations we used $N=4001$), and $\boldsymbol{\lambda}$ to be defined at the half grid points. We opt for a logarithmically distributed spatial grid because the torque is much stronger at the inner edge. In this way we can proceed with a leapfrog algorithm. The tracking of the evolution can be read in section $4.1$ of \citetalias{LOP02}.

\subsection{Results}
\label{sec:res_1D}

Henceforth in the whole paper we use $p=1/2$. We perform a first generic simulation with the following set of parameters: $(H_{\rm in}/R_{\rm in})=0.1$, $\alpha=0.05$, $\eta=0.25$ and $(a/R_{\rm in})=0.5$. We recall that $\eta=0.25$ corresponds to the case $M_1=M_2$. We make these choices because $H_{\mathrm{in}}/R_{\mathrm{in}}=0.1$ is the typical value for protostellar discs, when $\alpha=0.05$ we expect a low-viscosity behaviour, $M_1 = M_2$ gives a sizable torque, and finally $a/R_{\mathrm{in}}=0.5$ makes the approximation of a time independent gravitational potential reasonable. In the whole section we will then use: $x_{\rm in}=1$, $x_{\rm out}=90$ and $W_{\infty}=1$. The time unit will naturally be $\Omega_{\rm in}^{-1}$.

We consider a disc initially aligned with the binary plane in the inner parts, and misaligned in the outer parts. Thus as an initial condition we take $\lambda_x=0$ for $x\leq18$, $\lambda_x=\frac{1}{2}\{1+\sin{[\pi(x-20)/4]}\}$ for $18\leq x\leq22$, and $\lambda_x=1$ for $x\geq22$ \citepalias{LOP02}. $\lambda_y$ is initially set to $0$. In this way at $t=0$ we just have a tilt, with no twist. In other terms, $\gamma(x,t=0)=0$ at each radius.

\begin{figure}
\centering
\includegraphics[width=.9\columnwidth]{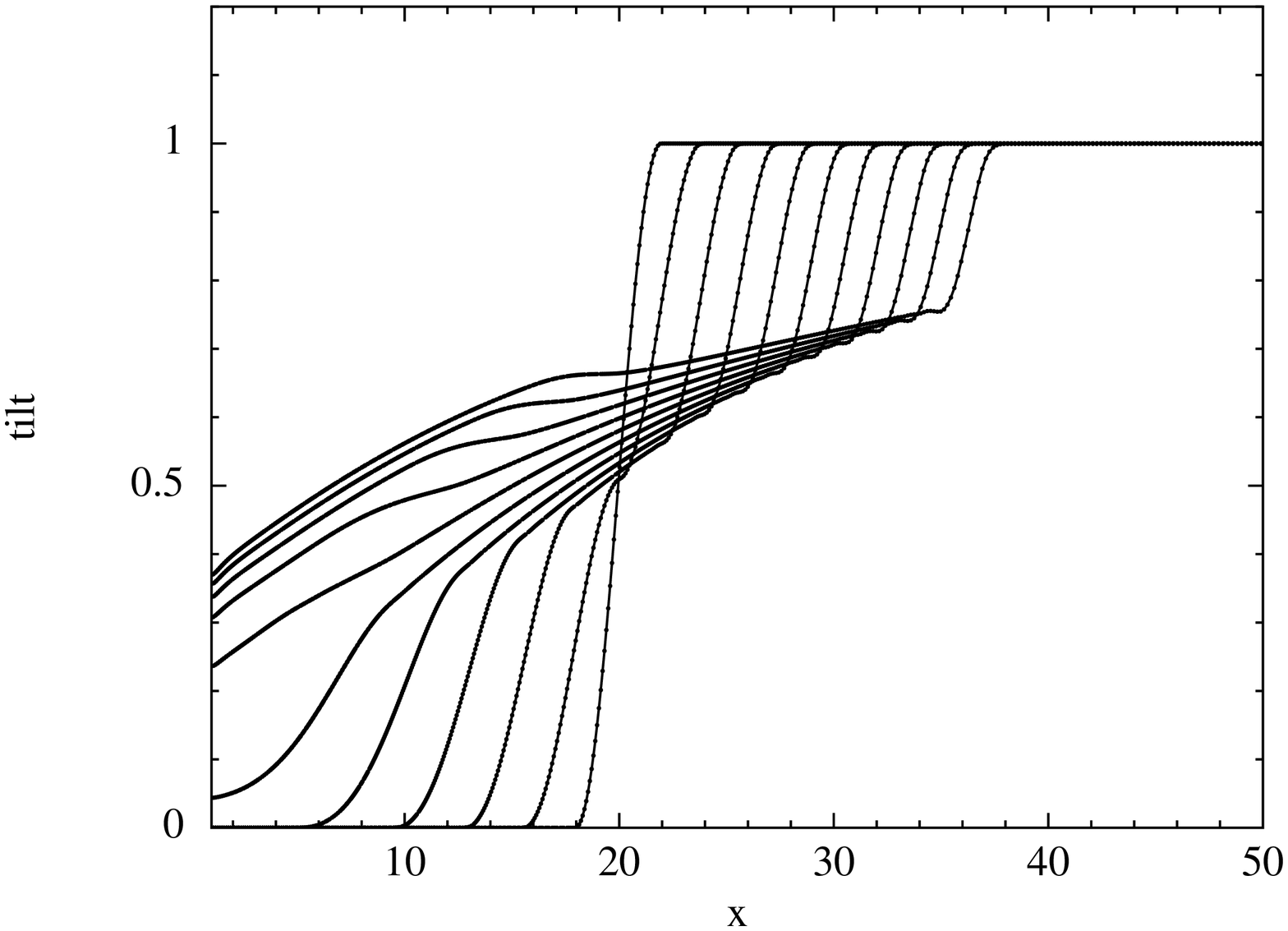}
\includegraphics[width=.9\columnwidth]{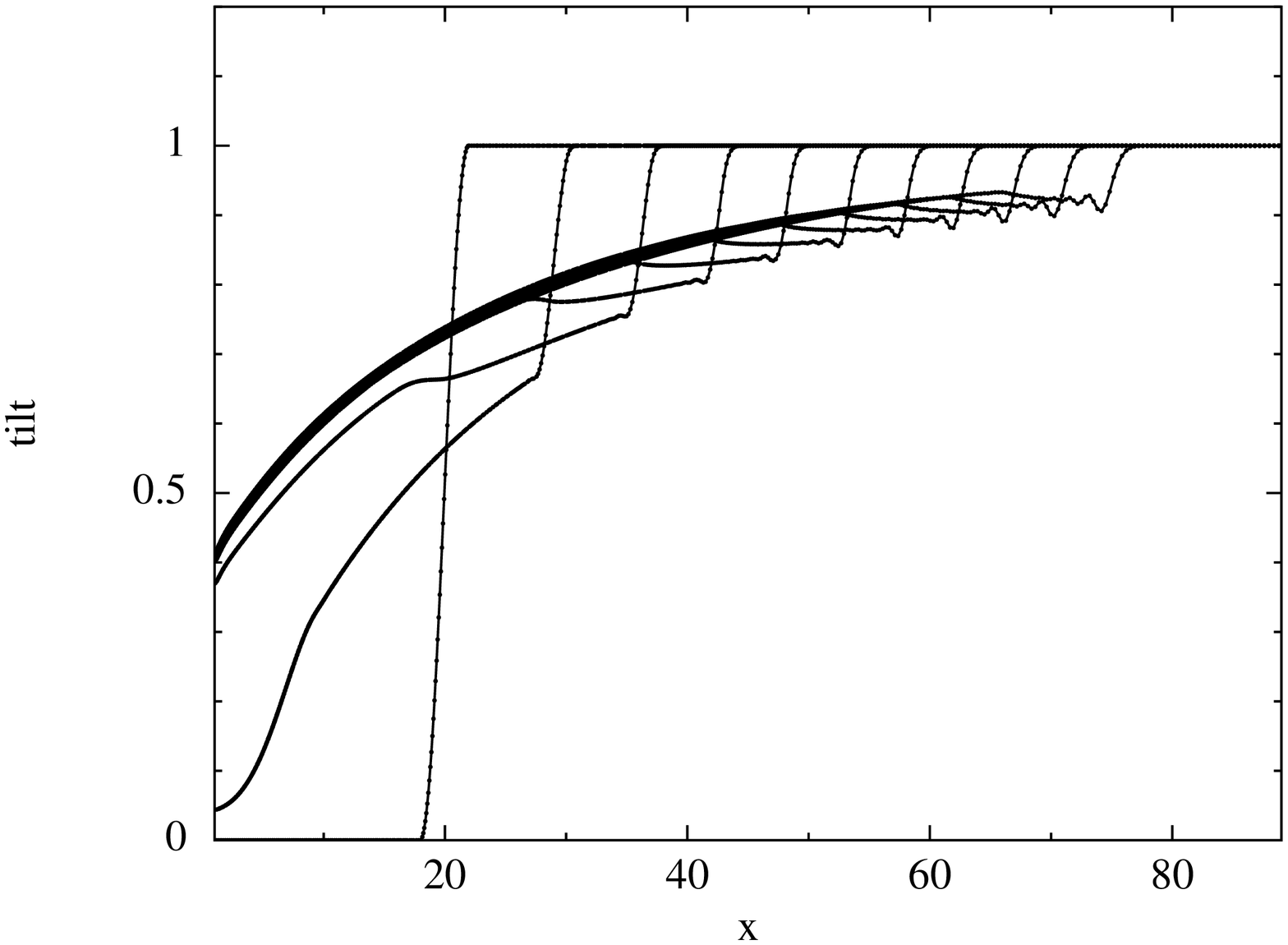}
\caption{Tilt evolution as a function of radius of an initially warped protostellar circumbinary disc in the linear low-viscosity regime, given the following set of parameters: $(H_{\rm in}/R_{\rm in})=0.1$, $\alpha=0.05$, $\eta=0.25$ and $(a/R_{\rm in})=0.5$. We recall that $\eta=0.25$ corresponds to $M_1=M_2$. Top panel: early evolution of the tilt, shown at 11 equally spaced times $t=0$, $400$, $800$,...,$4000$. Bottom panel: late evolution of the the tilt at 11 equally spaced times $t=0$, $2000$, $4000$,...,$20000$. As expected the discontinuity in the tilt initially propagates inwards and outwards in a wave-like fashion. The inward propagation of the tilt interacts with the strong external torque, and by the end of the simulation the steady state shape of the disc close to the binary is established. At $t=20000$ the initial outwardly propagating warp wave is approaching the outer edge of the grid, followed closely by the reflection of the initially inwardly propagating warp wave. By this time, the steady warped disc solution has been established over about half of the grid.}
\label{fig:wave_example_tstop4000}
\end{figure}

The evolution of the warp in this case is shown in Fig. \ref{fig:wave_example_tstop4000}. It is apparent that the discontinuity does propagate inwards and outwards as a bending wave. As the inwardly propagating wave reaches the inner edge, it bounces back and reacts to the strong external torque due to the binary, until it forms a stationary wave reaching a steady state on a sound crossing timescale. The outwardly travelling discontinuity keeps on propagating throughout the whole simulation. In the top panel of Fig. \ref{fig:wave_example_tstop4000} we report the tilt evolution, where the tilt is defined as $\sqrt{\lambda_x^2+\lambda_y^2}$ (normalised at $1$ at infinity) up to a computational time $t=4000$. The bottom panel of Fig. \ref{fig:wave_example_tstop4000} shows the same simulation up to $t=20000$. Note that the steady state has a tilt shape with the same features as the inviscid analytic solution: it is an evanescent wave, and the tilt does not tend to $0$ at the inner edge. These figures do not show the information about the twist, but we will analyse it later in this section. Finally, note that the waves propagate with a velocity $\approx c_{\rm s}/2$, as predicted by the dispersion relation in the low-viscosity case. From the figures we can observe small ripples propagating behind the outwardly travelling wave front. These are given by small numerical instabilities and are damped away by viscous interactions in almost a simulation time. We have verified that they can be reduced by increasing the spatial resolution. We have performed simulations with a set of initial conditions (e.g. an initially tilted untwisted disc), and the final state of the disc shape we obtain does not depend on them.

The warp in the presence of viscosity is much larger in this case with respect to the inviscid case (compare fig. \ref{fig:analytical} and \ref{fig:wave_example_tstop4000}). Also in this case, the warp amplitude agrees to order of magnitude with that predicted by \citet{foucart13} (for the same set of parameters, their equation 16 would predict $\Delta\beta/\beta_{\infty}\approx 0.4$, and we observe $\Delta\beta/\beta_{\infty}\approx 0.6$, where $\beta_{\infty}$ is the tilt at large radii). 

\begin{figure}
\centering
\includegraphics[width=.9\columnwidth]{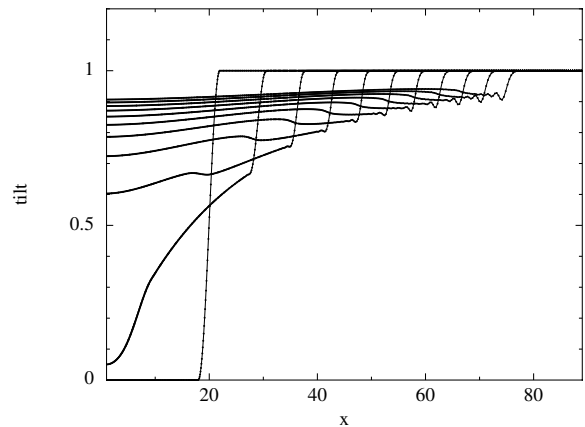}
\caption{Tilt evolution as a function of radius at 11 equally spaced times $t=0$, $2000$, $4000$,...,$20000$ with $(H_{\rm in}/R_{\rm in})=0.1$, $\alpha=0.05$, $\eta=0.25$ and $(a/R_{\rm in})=0.1$. At equal viscosity, when the external torque is less significant the disc tends to stay misaligned, since the torque is not sufficiently strong to induce a warp against internal pressure stresses.}
\label{fig:gravity}
\end{figure}

\begin{figure}
\begin{center}
\includegraphics[width=.9\columnwidth]{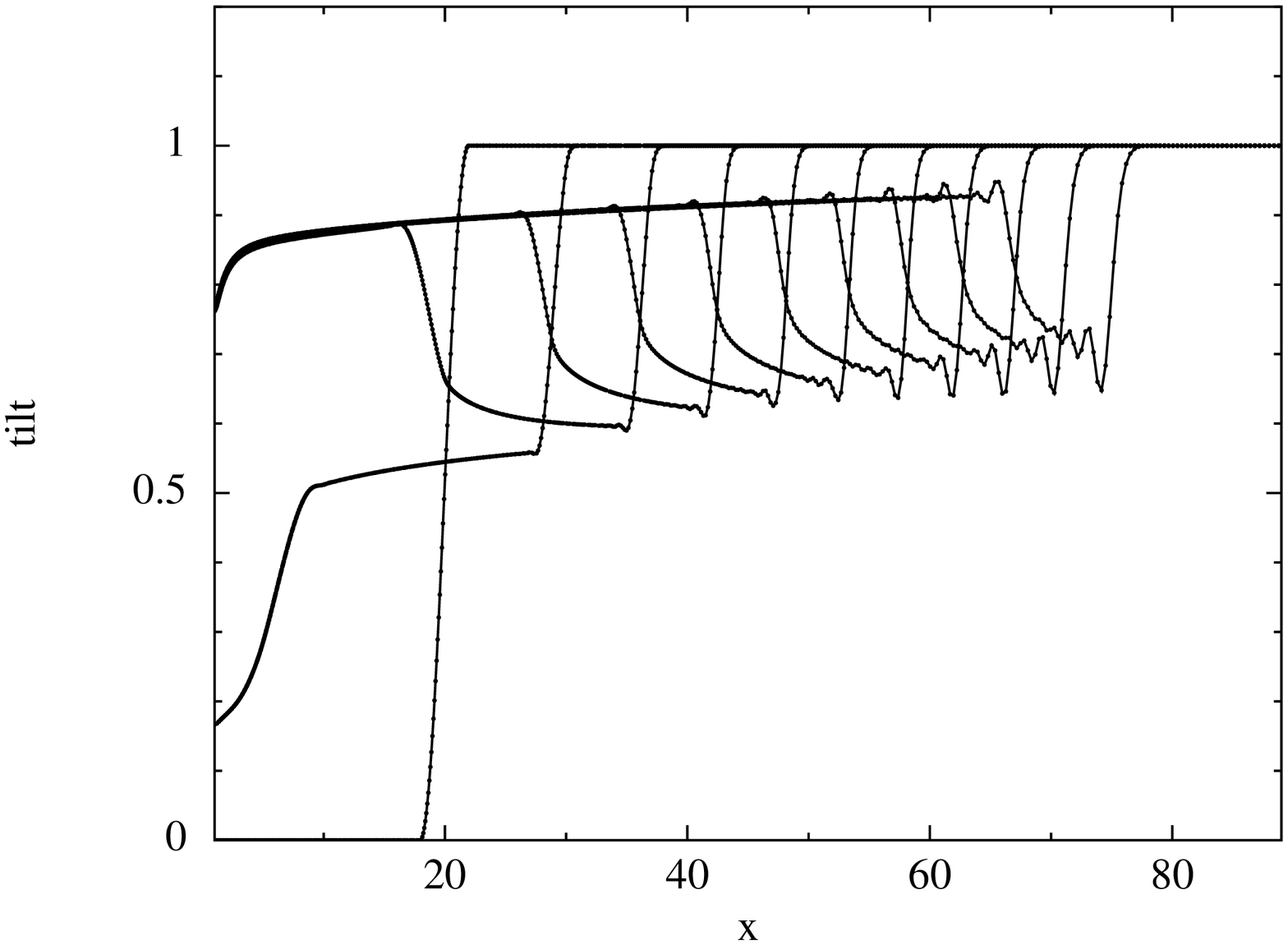}
\includegraphics[width=.9\columnwidth]{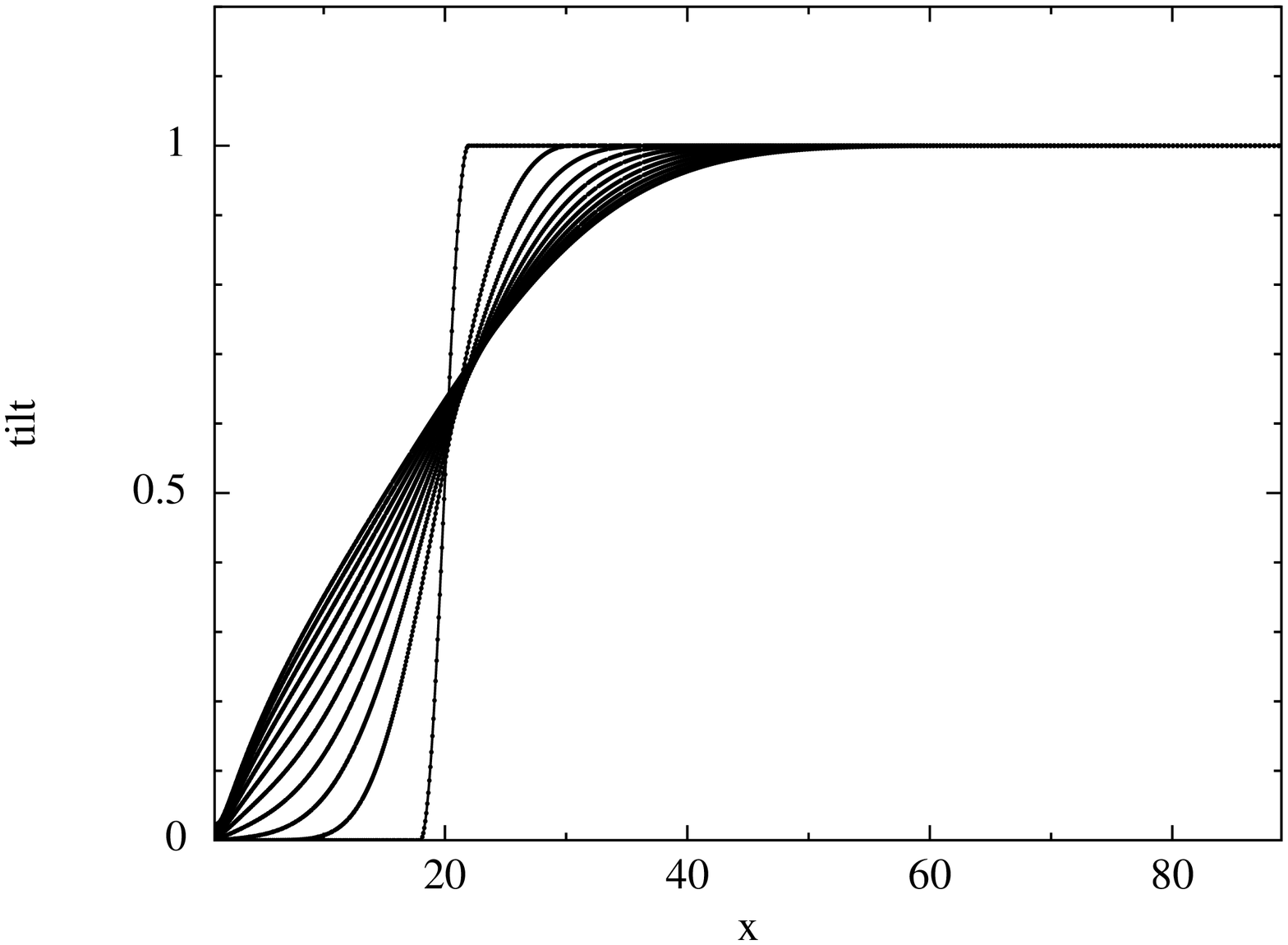}
\caption{Top panel: tilt evolution for an initially warped disc with $(H_{\rm in}/R_{\rm in})=0.1$, $\alpha=0.01$, $\eta=0.25$ and $(a/R_{\rm in})=0.5$ at 11 equally spaced times $t=0$, $2000$, $4000$,...,$20000$. Bottom panel: tilt evolution for the same disc with $\alpha=0.6$. In the less viscous disc the warp propagates at a half of the sound speed, and the shape of the tilt tends to a final steady state which is comparable to the analytic solution for completely inviscid discs. In the figure reported in the bottom panel we are in a diffusive regime ($\alpha>H/R$); the diffusive behaviour is apparent.}
\label{fig:visc}
\end{center}
\end{figure}

We have then explored the parameters space by modifying the two parameters governing the physics of the system: $\chi$ and $\alpha/(H_{\rm in}/R_{\rm in})$. By modifying $\chi$, we change the physical parameters of the binary system, either the distance between the two stars or the ratio between the two masses. Fig. \ref{fig:gravity} shows the tilt evolution when $a/R_{\rm in}=0.1$, $\eta=0.25$, $H_{\rm in}/R_{\rm in}=0.1$ and $\alpha=0.05$. The disc edge lies further away from the binary (the inner radius is larger than the tidal truncation radius). This choice is equivalent to reducing the binary mass ratio and keeping the inner radius fixed. In this case, the external torque due to the non spherical symmetry of the potential is reduced, and therefore the disc tends to stay misaligned with respect to the binary plane (see also Fig. \ref{fig:analytical}). In Fig. \ref{fig:visc} we portray two simulations with different values of viscosity. The top panel has $\alpha=0.01$, the bottom one $\alpha=0.6$, while the other parameters are $a/R_{\rm in}=0.5$, $\eta=0.25$ and $H_{\rm in}/R_{\rm in}=0.1$. The second one lies in the diffusive regime ($H/R<\alpha<1$), and it shows its characteristic behaviour. We can note two relevant differences between the two panels. First, in the diffusive simulation the disc tends to align with the binary plane in the inner parts, as already illustrated in previous works \citep{lodato_pringle06, lodato_pringle07}.  Secondly, in this case the evolution of the shape of the disc slows down significantly, since the dispersion relation does heavily depend on $\alpha$.

\begin{figure}
\begin{center}
\includegraphics[width=.9\columnwidth]{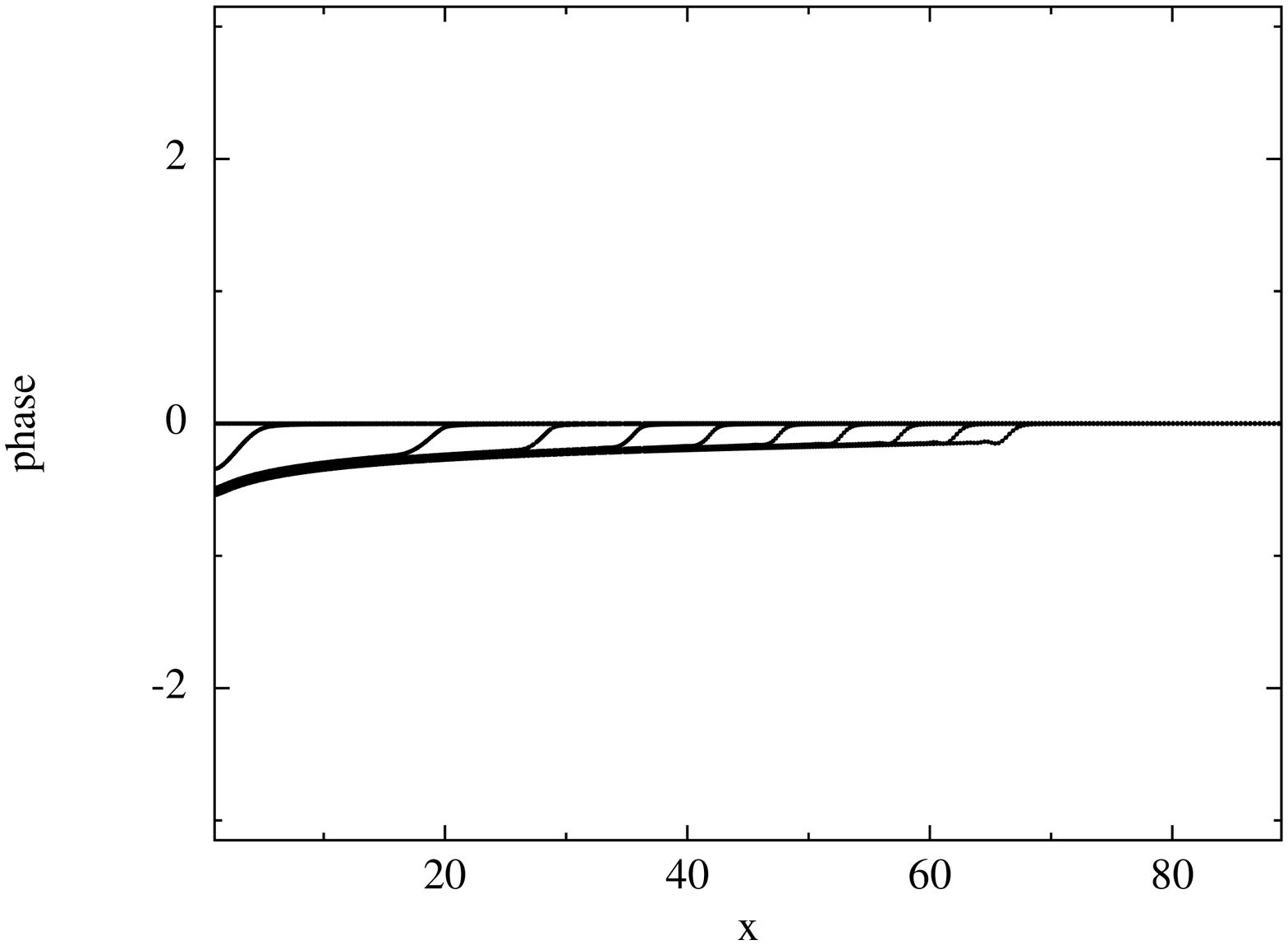}
\includegraphics[width=.9\columnwidth]{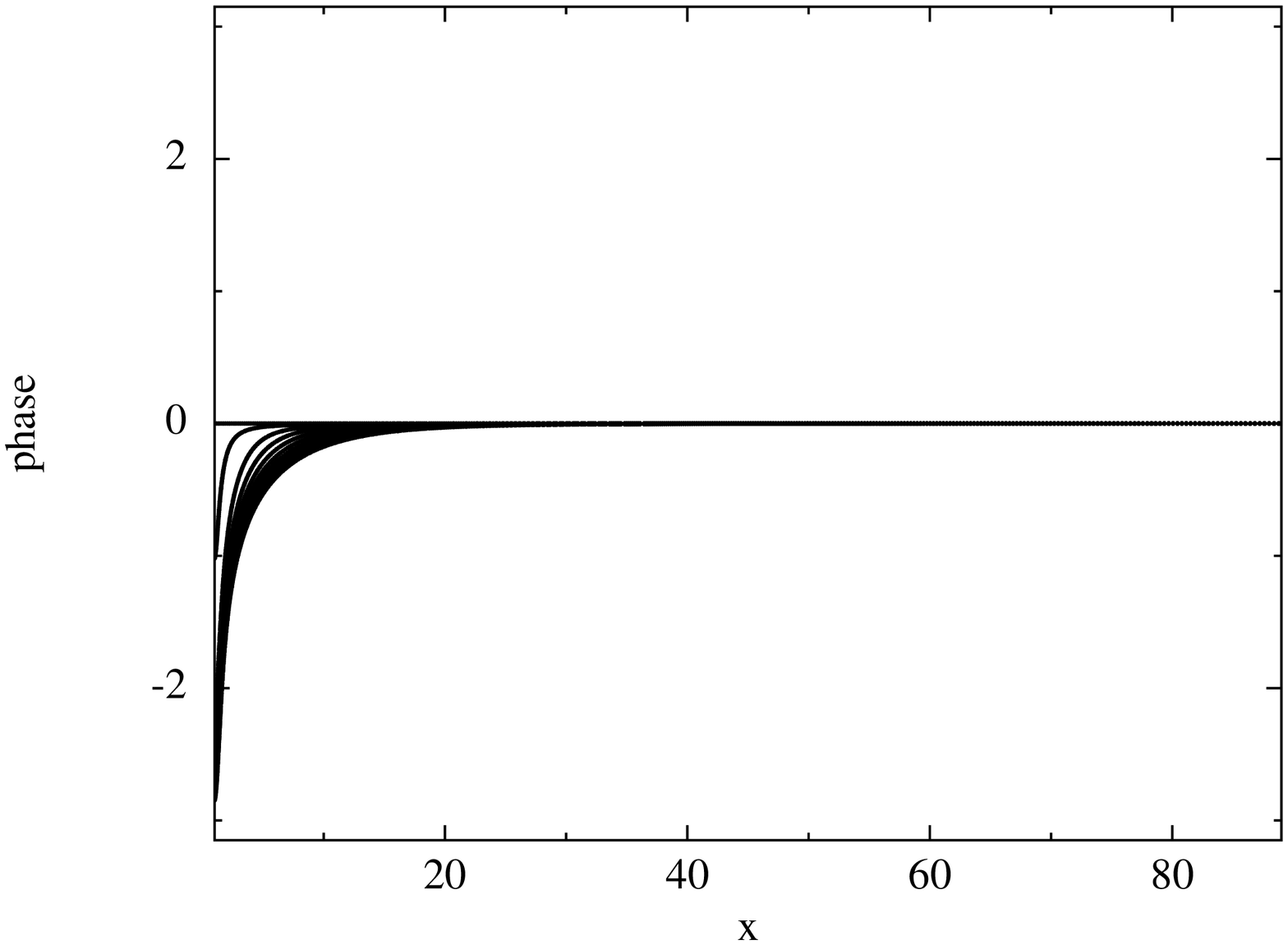}
\caption{Phase evolution of the same discs portrayed in Fig. \ref{fig:visc} (top panel: $\alpha=0.01$; bottom panel: $\alpha=0.6$). The twist of the disc correlates with the viscosity: the more viscous the disc is, the more twisted it gets. In the bottom panel the twisting is so strong that $\Delta\gamma/\Delta R\approx2\pi$.}
\label{fig:visc_phase}
\end{center}
\end{figure}

So far we have not specified how the final solutions differ from one another with respect to the phase. In Fig. \ref{fig:visc_phase} we illustrate the phase (as a function of radius) for the two cases reported in Fig. \ref{fig:visc}. We can observe that both discs are twisted in the inner regions. Moreover, the magnitude of the twist increases as the viscosity increases. The twisting of the disc is therefore strongly correlated to viscosity (in fact viscosity is the only physical quantity that can induce shear forces in this case). The limit case is $\alpha=0$; in section \ref{sec:inviscid} we will see that inviscid discs present no twist at all for $t\rightarrow\infty$, independently of the initial condition.

We can compare these results with the ones obtained by \citet{foucart13}. We agree with the fact that the steady state solutions depend on two parameters only: $\alpha/(H_{\rm in}/R_{\rm in})$ and $\chi$, as they show in their equations 15, 16 and 20 when they estimate the amplitude of both the warping and the twisting. Moreover, we confirm that equation 20 gives a right estimate of the warping of the disc when the disc is (nearly) inviscid (see Fig. \ref{fig:analytical}). When $\alpha/(H_{\rm in}/R_{\rm in})$ is not negligible, we confirm that the warping is dominated by the term given by their equation 16.

\subsection{The inviscid case}
\label{sec:inviscid}

\begin{figure}
\begin{center}
\includegraphics[width=.9\columnwidth]{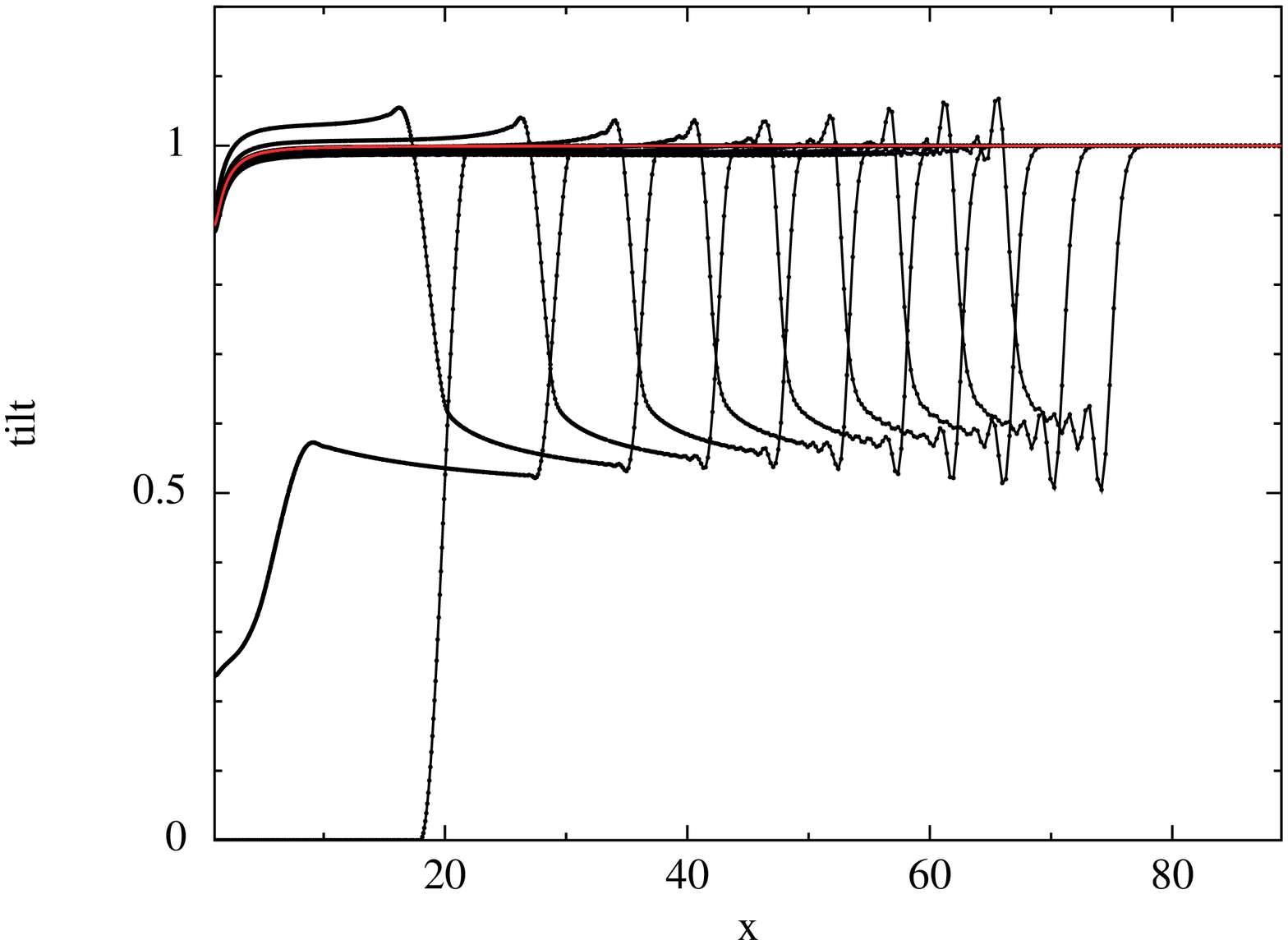}
\includegraphics[width=.9\columnwidth]{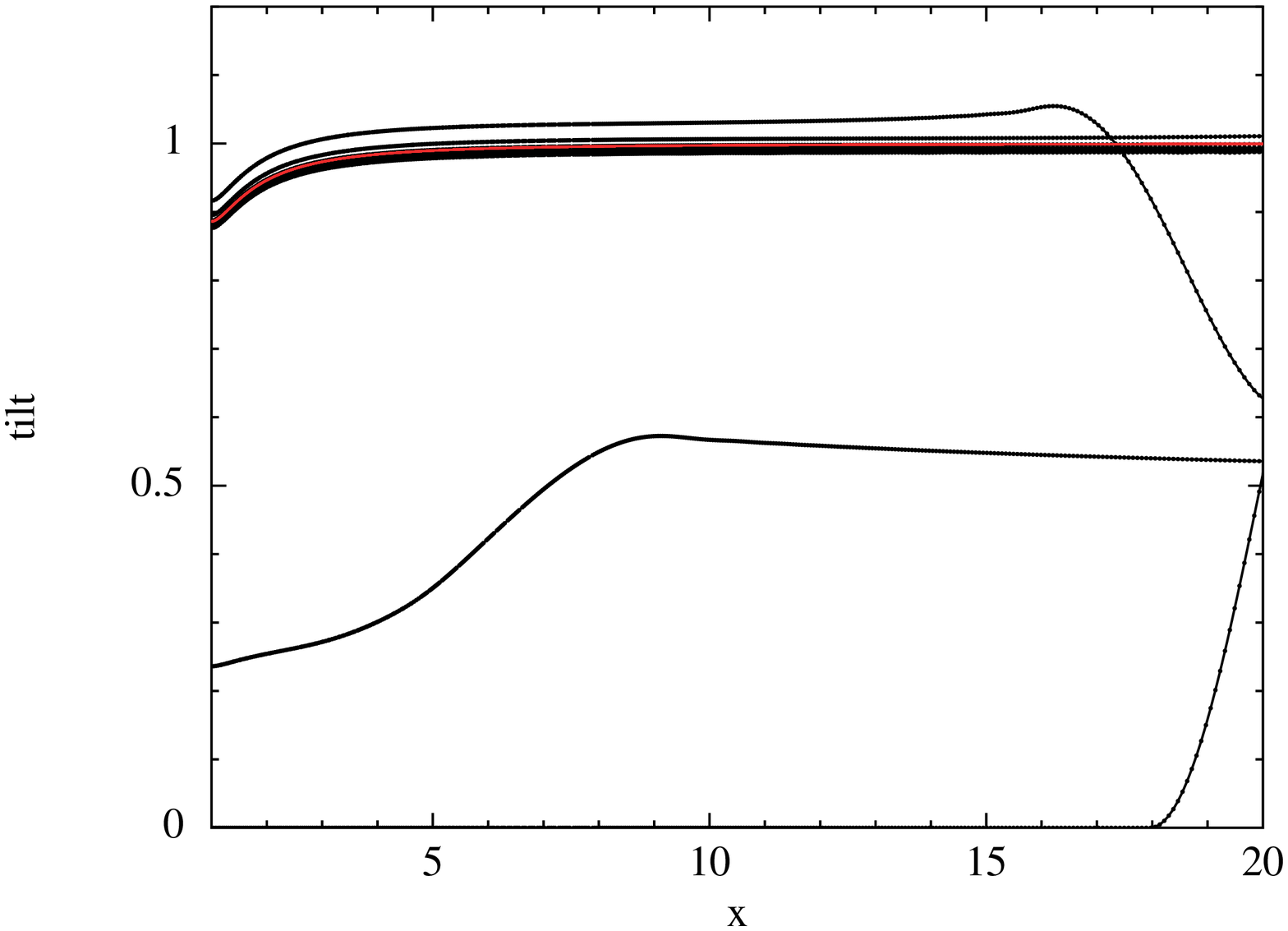}
\caption{Top panel: tilt evolution of an inviscid initially warped disc with $(H_{\rm in}/R_{\rm in})=0.1$, $\eta=0.25$ and $(a/R_{\rm in})=0.5$ at 11 equally spaced times $t=0$, $2000$, $4000$,...,$20000$. The red line depicts the analytic solution of the steady state's shape. Bottom panel: blow up of the inner region.}
\label{fig:alpha_0}
\end{center}
\end{figure}

\begin{figure}
\centering
\includegraphics[width=.9\columnwidth]{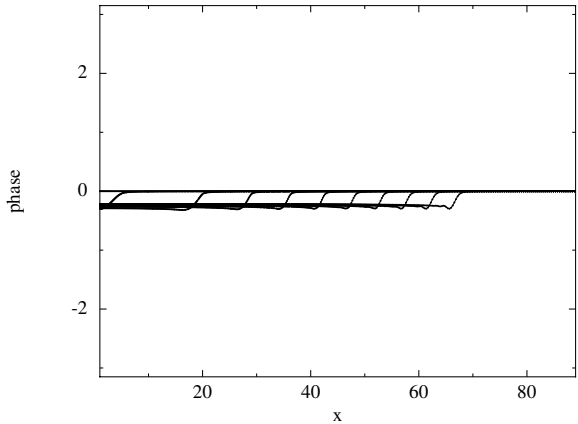}
\caption{Phase evolution of the disc depicted in Fig. \ref{fig:alpha_0}. The initial condition is $\gamma(R)=0$. The disc tends to a steady state with $\gamma=$const. The disc rearranges towards the analytic solution in the tilt; in order to do that, it initially gains a twist, that then propagates outwards while the disc reaches an untwisted shape. This fact confirms that an inviscid disc tends to a steady state that is well described by the analytic solution both in the tilt shape and in the untwisted behaviour.}
\label{fig:alpha_0_phase}
\end{figure}

We now compare the analytic solution for the steady state shape (section \ref{sec:analyt}) with the results of the 1D time-dependent simulations. We recall that we deduced the analytic solution for inviscid discs only, therefore we have to set $\alpha=0$ in order to be able to compare our results. We consider an initially warped disc with $R_{\mathrm{in}}=2a$, $H_{\mathrm{in}}/R_{\mathrm{in}}=0.1$ and $\eta=0.25$. Fig. \ref{fig:alpha_0} shows the comparison with the analytic solution (highlighted with the red line). For this simulations we have used $N=4001$ (number of grid points) in order to reduce the ripples due to low resolution. The agreement is good: the time-dependent calculation leads to a steady state that is well described by the analytic solution. The disc rearranges in such a way that it reaches a steady state with a constant phase. As predicted by the analytic model, time-dependent inviscid simulations tend to an untwisted steady state, which is uniformly rotated with respect to the initial condition (see Fig. \ref{fig:alpha_0_phase}). Such rotation is needed because in order to reach the steady warped configuration, the disc has to mix the $x$- and $y$- components of ${\bf l}$, thus producing a transient twist that eventually settles in an untwisted, but rotated, configuration. These results have been confirmed by using different initial conditions. Moreover, we ran a simulation with the analytic solution as initial condition. We see little evolution in this case, mostly driven by small numerical noise.

\section{Full 3D simulations}
\label{sec:res}

Hitherto in order to describe warp propagation in protostellar circumbinary discs, we have made many approximations. We recall them here. First of all, we have considered the gravitational potential generated by the two central stars as time independent. In order to do so, we assumed that the inner edge of the disc was quite far from the binary, $r_1/R_{\mathrm{in}} < 1$ and $r_2/R_{\mathrm{in}} < 1$. Secondly, in the thin-disc approximation, we have considered $\partial_t \Sigma = 0$, because the viscous time scale is much longer than the sound crossing time, which is the warp evolution time in the wave-like regime. Then, we have considered the linear (small amplitude) waves propagating with the single $m=1$ mode.

3D hydrodynamical simulations allow us to model our physical systems without making any of the assumptions listed above. First of all, we can implement the rotation of the two central stars directly without estimating the time-independent contribution of their gravitational potential. The particles describing the disc flow will be subject to the full potential generated by the binary. Secondly, we do not have to make any assumption about the warp dynamics, and especially the assumption of linear perturbations. By comparing the results of the 3D simulations with the results obtained by the 1D code, we will be able to check whether our assumptions were reasonable, and in which parameters range we can consider them valid.

\subsection{SPH and viscosity}
\label{sec:SPH}

We perform our 3D simulations by using a smoothed particle hydrodynamics (SPH) code \citep[see][for a recent review]{price12}. We have used the \textsc{phantom} code by Daniel Price \citep[see e.g.][]{lodato_price10, price_fed10}, which has been shown to perform well in dealing with warp propagation in the diffusive (non self-gravitating) regime \citep[see][for two recent applications]{lodato_price10,nixon_al12}. As already mentioned in the introduction, SPH codes have already been used to simulate warp propagation in the bending-wave (thick disc) regime, but with very low resolution \citep{nelson_pap99,nelson_pap00}.

In this section we shall not go through SPH theory. We will just describe how we implemented an isotropic viscosity in the code. We know that SPH codes implement an artificial viscosity in order to spread discontinuities over a few smoothing lengths. It has been known for some time \citep{murray96} that the artificial terms in SPH can be understood straightforwardly as numerical representations of second derivatives of the velocity, and this makes viscosity act also when there is a purely shear flow, as in an accretion disc. The $\alpha$ parameter for the shear viscosity is related to the artificial one ($\alpha_{\rm art}$) by:
\begin{equation}
\alpha_{\rm SS}=\frac{1}{10}\alpha_{\rm art} \frac{\bar{h}}{H},
\label{eq:visc_sph}
\end{equation}
where $\bar{h}$ is the averaged smoothing length at radius $R$ and $H$ the scale-height of the disc at the same radius. \citet{lodato_price10} showed excellent agreement between this relation and the outcome of their simulations. The notation $\alpha_{\rm SS}$ is used to discriminate between the directly implemented physical viscosity (see below) and the physical viscosity due to the artificial one. Thus, we could simulate the physical viscosity of our discs by using the artificial one. However, in order to keep a uniform value for $\alpha_{\rm SS}$, we would need a constant ratio $\bar{h}/H$ in the disc. In the case of power-law density and sound speed profiles, this requires the two exponents to be related by $p+2q=3$ (e.g. $p=1.5$, $q=0.75$). Our choice for these two parameters is different though, since we preferred to use $p=0.5$ in order to compare our 1D code with \citetalias{LOP02}. To mimic viscosity, in this work we used an alternative formulation proposed by \citet{flebbe94}, where we evaluate directly the stress tensor in the Navier-Stokes equation. We preferred this formulation to the similar one by \citet{espanol_revenga07} because the former has been well tested by \citet{lodato_price10}, and the latter does not conserve angular momentum. By this method, we have direct control of the viscous terms, since we set the shear viscosity by hand:
\begin{equation}
\nu=\alpha c_{\rm s}^2(R)/\Omega(R),
\end{equation}
where $\alpha$ is the chosen value of the viscosity parameter, and the profiles of the sound speed $c_{\rm s}$ and of the (Keplerian) angular velocity $\Omega$ are prescribed functions of $R$. Moreover, we can set the bulk viscosity to $0$. In this work, however, we still keep a small amount of artificial viscosity in order to correctly dissipate shocks if they are present and prevent particle interpenetration using the \citet{morris_mon97} switch. Moreover, when the physical $\alpha$ is set to $0$ in order to simulate an inviscid motion, we will need an artificial viscosity in order to prevent chaotic motions of SPH particles \citep{price_fed10} that would increase the effective viscosity instead of reducing it. In the simulations presented in this section, we have used $\alpha_{\rm art,max}=0.5$ and $\alpha_{\rm art,min}=0.01$ (unless specified otherwise), where the value of $\alpha_{\rm art}$ between this two values is estimated via the Morris and Monaghan switch. The \citet{von50} $\beta_{\rm art}$ parameter has been set equal to $2$.

Finally, note that in our simulations we did not compute the energy equation. We adopted a locally isothermal equation of state (set by the $q$ parameter assigned to the local sound-speed).

\begin{figure*}
\begin{center}
\includegraphics[width=.9\columnwidth]{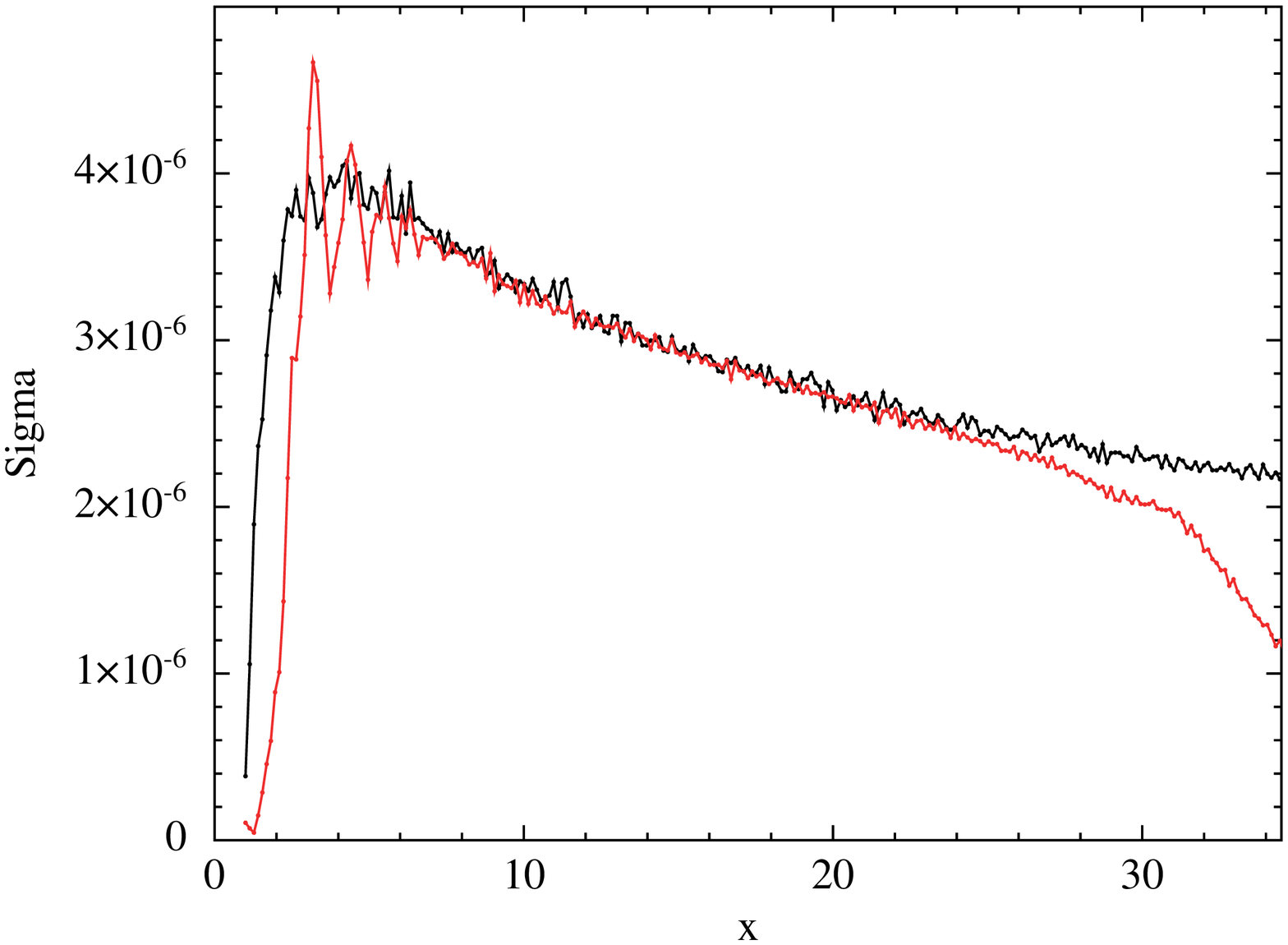}
\includegraphics[width=.9\columnwidth]{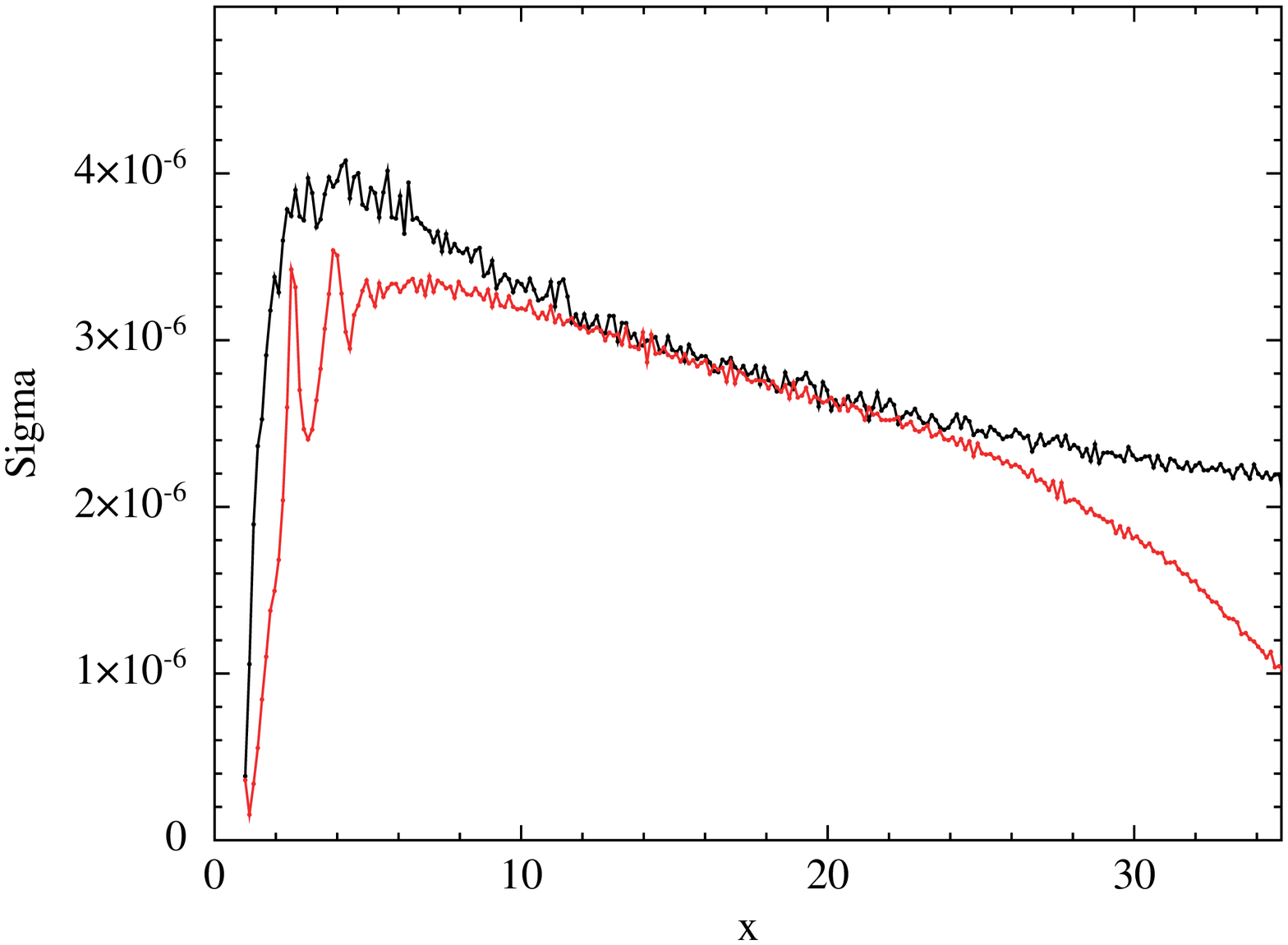}
\end{center}
\caption{Left panel: evolution of $\Sigma$ for an initially untwisted tilted disc with $\alpha = 0.05$, $H_{\rm in}/R_{\rm in}=0.1$, $\eta=0.25$, $R_{\rm in}=a$ and $\beta_{\infty}=5^\circ$. Right panel: evolution of $\Sigma$ for the same setup but with $\alpha = 0.2$. The black lines refer to $t = 0$, while the red lines to $t = 4000$, which is the end of both simulations.}
\label{fig:sigma_evol}
\end{figure*}

\begin{figure*}
\begin{minipage}{180mm}
\begin{center}
\includegraphics[width=.95\columnwidth]{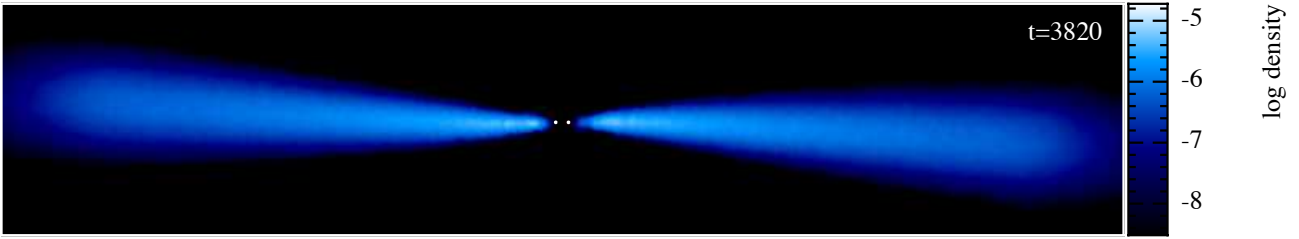}
\end{center}
\end{minipage}
\caption{Cross section of the inner regions of the disc in the SPH calculations at a resolution of 1 million particles with $\alpha = 0.05$, $H_{\rm in}/R_{\rm in}=0.1$, $\eta=0.25$ and $\beta_{\infty} = 5^\circ $ at $t = 3820$.  The two white points represent the two central stars. The colour scale indicates density in code units.}
\end{figure*}

\begin{figure*}
\begin{center}
\includegraphics[width=.9\columnwidth]{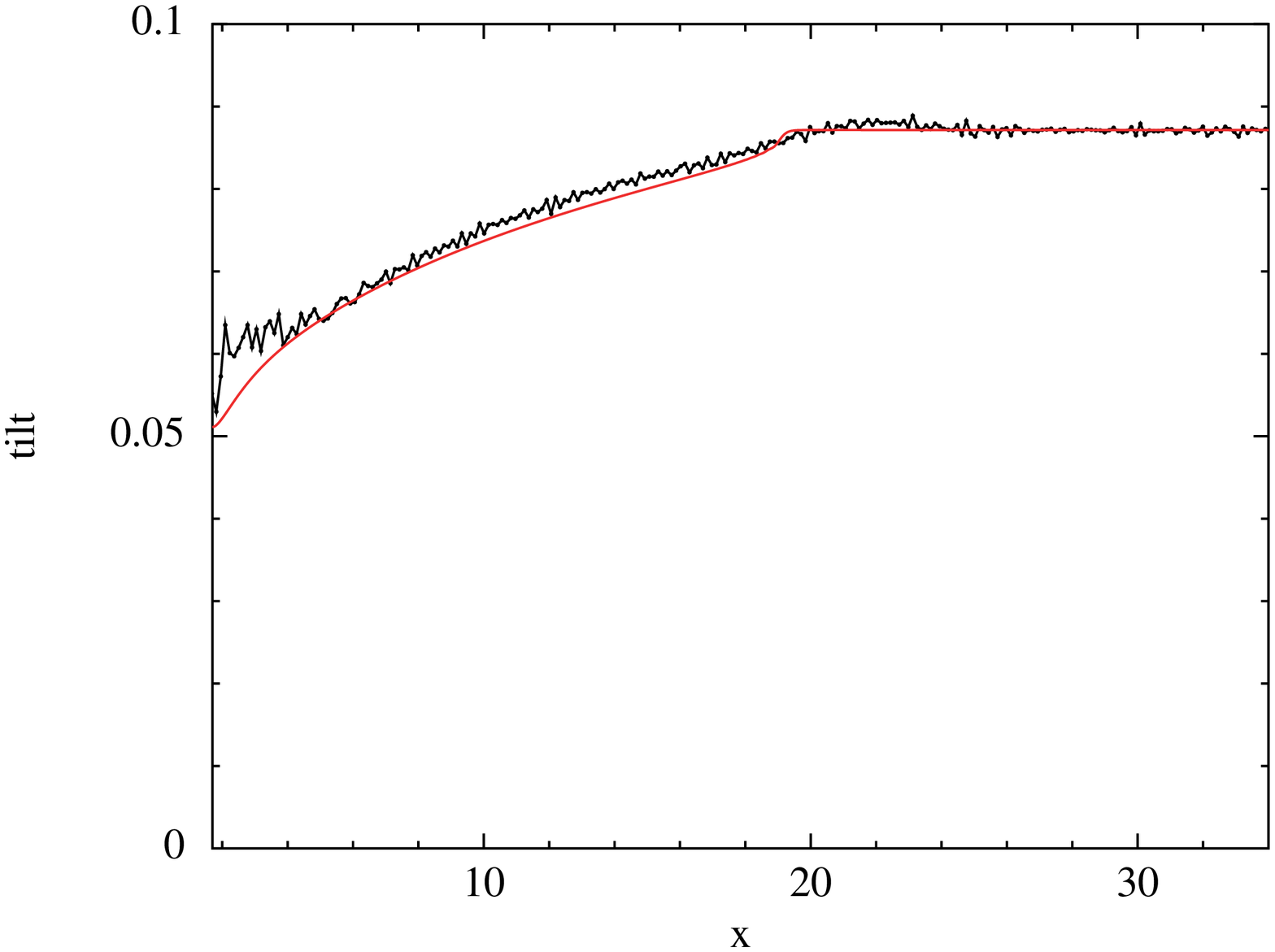}
\includegraphics[width=.9\columnwidth]{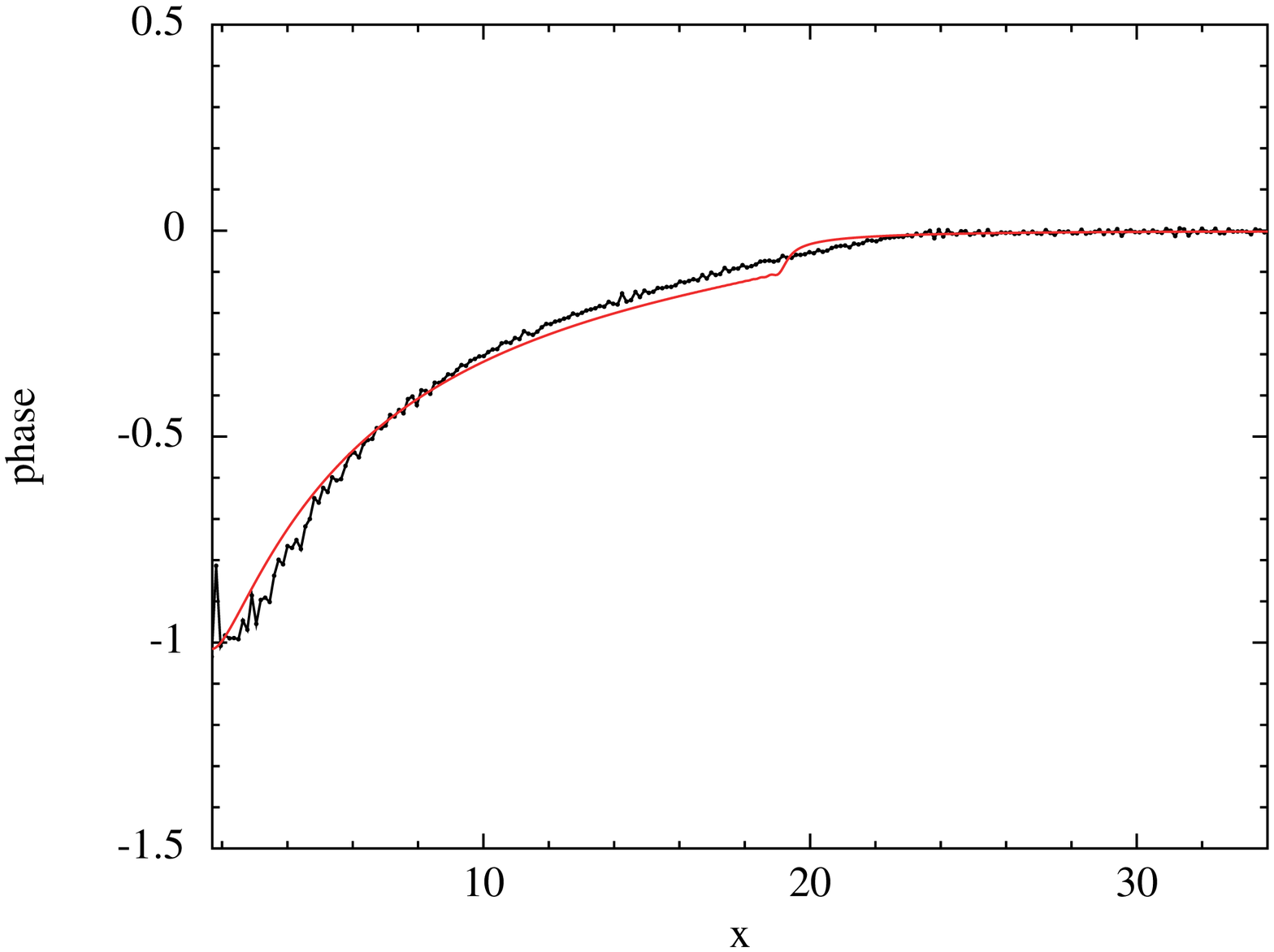}
\end{center}
\caption{Tilt and phase evolution of an initially untwisted tilted disc, with $\eta=0.25$, $(H_{\rm in}/R_{\rm in})=0.1$, $\alpha = 0.05$, $\beta_{\infty} = 5^\circ$ and $N = 1$ million at $t=2000$. The red line depicts the corresponding 1D simulation, where the inner edge of the disc has been set equal to $R_{\rm t}=1.7 a$. The agreement between the SPH simulations and the 1D results is very high, both for the tilt and for the phase of the warp. The propagation velocities of the 3D and the 1D simulations coincide.}
\label{fig:res_005}
\end{figure*}

\subsection{Numerical setup and initial conditions}

We model the two central stars as sink particles. i.e. non-gaseous particles with appropriate boundary conditions \citep{bate95}. We assign an accretion radius to each of the two sink particles, i.e. the radius within which we can consider the fluid as accreted onto the star. This feature allows us not to follow the dynamics of gas particles too close to the binary system, since it would be computationally expensive. We assign an accretion radius of $0.05 R_{\rm in}$ to each star.

We have run simulations with different resolutions. We go from $N=10^5$ to $N=2\cdot10^6$, where $N$ is the number of particles used. Once we have assigned the values of the physical parameters of the system (see below), we distribute the particles so that the disc attains a prescribed initial density profile. We assign each particle a radius dependent sound-speed $c_{\rm s}$, where $c_{\rm s}\propto R^{-q}$ as usual. We then distribute the particles in the vertical direction, in order for the density to have a Gaussian profile in the $z$-direction. The scale-height is set to $H=c_{\rm s}/\Omega$. Finally, the initial angular velocity is assigned by taking into account pressure contributions:

\begin{equation}
v_{\phi}=v_{\rm K}\left[1-(p+2q)\left(\frac{c_{\rm s}}{v_{\rm K}}\right)^2\right]^{1/2},
\end{equation}
where $v_{\rm K}=\sqrt{GM/R}$ is the usual Keplerian velocity.

As initial condition, we implement a surface density equal to:

\begin{equation}
\label{eq:sigma_in}
\Sigma(R)=\Sigma_0 R^{-p} \left( 1-\sqrt{\frac{R_{\mathrm{in}}}{R}} \right),
\end{equation}
where $p$ is the usual coefficient introduced in section \ref{sec:analyt}. Note that this is slightly different from the pure power law used above in section \ref{sec:analyt}. We use the above relation because otherwise the inner gas would be pushed inwardly by the strong pressure gradient at the inner edge. Moreover, with this setup it is reasonable to consider the $x$ and $y$ components of the torque equal to $0$. Note that $\Sigma_0$ does not play a key role in our simulations, as long as $M_d \ll M_1+M_2$, where $M_d$ is the mass of the disc. In fact, self-gravity is not implemented in the SPH code. However, sink particles do feel the gravitational force generated by the gas particles. Since $M_d \ll M_1+M_2$, the back-reaction of the disc on the angular momentum of the binary is negligible ($M_d=0.01M$ in all the simulations). We do not impose any condition at the outer edge. We just consider wide discs, in such a way that $t_{\nu} = R_{\rm out}^2/\nu \gg t_{\mathrm{sound}}$, where $t_{\rm sound}$ is the sound crossing time, so that the external boundary condition does not affect the evolution in the inner regions, which are the ones we are interested in. However, we do not simulate too wide discs because most of the mass lies in the outer regions ($M(R)\propto R^{2-q}$), but we need a high resolution at the very inner edge where the external torque is stronger. The dynamic range of our simulations is $R_{\rm out}/R_{\rm in}=35$.

The other parameters have been set to: $p=0.5$, $q=0.75$, $H_{\rm in}/R_{\rm in}=0.1$, $\eta=0.25$ and $R_{\rm in}=a$. This set of parameters is slightly different from our standard runs of section \ref{sec:1D}. In particular, here we consider a disc that is initially closer to the central binary. We made this choice in order to have a more prominent warp. Otherwise, since the resolution at small radii is relatively poor (because $\Sigma\rightarrow 0$), the features of the warp would be masked by the low signal to noise ratio. Note however that the inner radius of the disc will be pushed further from the central binary due to tidal forcing. We will discuss this issue in detail in section \ref{subsec:results}. The time variable $t$ is expressed in terms of $\Omega_{\mathrm{in}}^{-1}$ in the whole section.

As initial condition we used an untwisted disc uniformly tilted with respect to the binary plane. Therefore, at $t=0$, $\beta(R)={\rm const}$ and $\gamma(R)=0$. The initial inclination angle and the physical viscosity will be specified later while presenting the results. The simulations are run in the same dimensionless units as in section \ref{sec:analyt}.

\begin{figure}
\begin{center}
\includegraphics[width=.9\columnwidth]{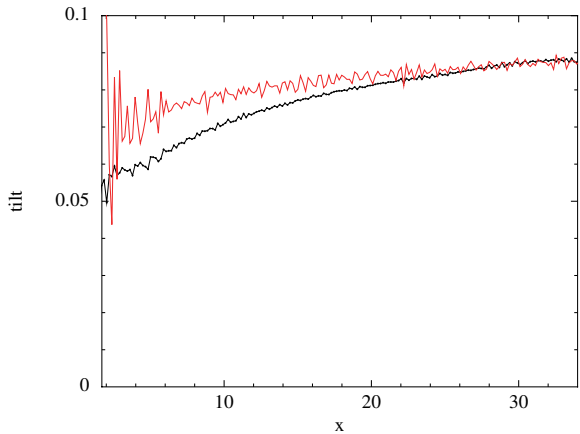}
\end{center}
\caption{Tilt of two differently resolved discs, with $\alpha=0.05$ and $\beta_{\infty}=5^\circ$, at $t=4000$. The black line portrays a disc with $N=1$ million particles, and the red line a disc with $N=100000$ particles. For less well resolved discs, the tidal torques at the inner disc edge are less effective, leading to a smaller warp. The number of shells to compute the averaged quantities is $200$ in order to reduce the scatter in the inner regions.}
\label{fig:res_resolution}
\end{figure}

\subsection{Results}
\label{subsec:results}

In this section we compare the results of the 3D SPH simulations with the ones obtained by the 1D code. New 1D simulations are reported here, with a surface density profile equal to the one reported in equation \ref{eq:sigma_in}. In order to compare the results we had to compute azimuthally averaged disc quantities of the the SPH simulations in a number of thin shells. The procedure is the one described in section $3.2.6$ of \citet{lodato_price10}. The number of shells has been set to $300$.

Before comparing the results, we have verified that $\partial_t \Sigma\approx0$ in the SPH simulations. We looked at the evolution of $\Sigma$ in a simulation time ($t_{\mathrm{stop}} = 4000$) for two cases: $\alpha = 0.05$ and $\alpha = 0.2$. Note that with our setup $t_{\nu}\approx10^5(0.2/\alpha)\gg t_{\rm stop}$. For these two simulations $N = 1$ million. In Fig. \ref{fig:sigma_evol} we show $\Sigma$ at the beginning of the two simulations and at their very end. From the results illustrated in the figure we can conclude that the assumption $\partial_t \Sigma \approx 0$ is satisfied at least in the bulk of the disc, even in the diffusive regime ($\alpha=0.2$). However, we can make the following observations.

Firstly, at the outer edge the surface density profile smooths towards a continuous configuration. The initial condition presents a discontinuity at the outer radius, which is damped out quite quickly by pressure forces. Secondly, and more importantly, in both cases we observe an evolution at the inner edge: the inner radius is pushed further from the binary by tidal forcing, as we expect since the initial inner radius is smaller than the tidal truncation one. From \citet{art_lubow94} we expect it to be at a radius $R_{\rm t}\approx 1.7 a$ when $\eta = 0.25$. Indeed, in section \ref{sec:analyt} we had set this value equal to $2a$ for simplicity. Note that the location of the inner disc edge has a strong effect on the warp, given the strong radial dependence of the binary torques. In the two cases shown in Fig. \ref{fig:sigma_evol} $R_{\rm t}$ lies around $1.5-1.9 a$, in good agreement with the predicted $1.7 a$.

Finally, we have verified that the binary angular momentum variations are negligible over the simulation time.

\begin{figure*}
\begin{center}
\includegraphics[width=.9\columnwidth]{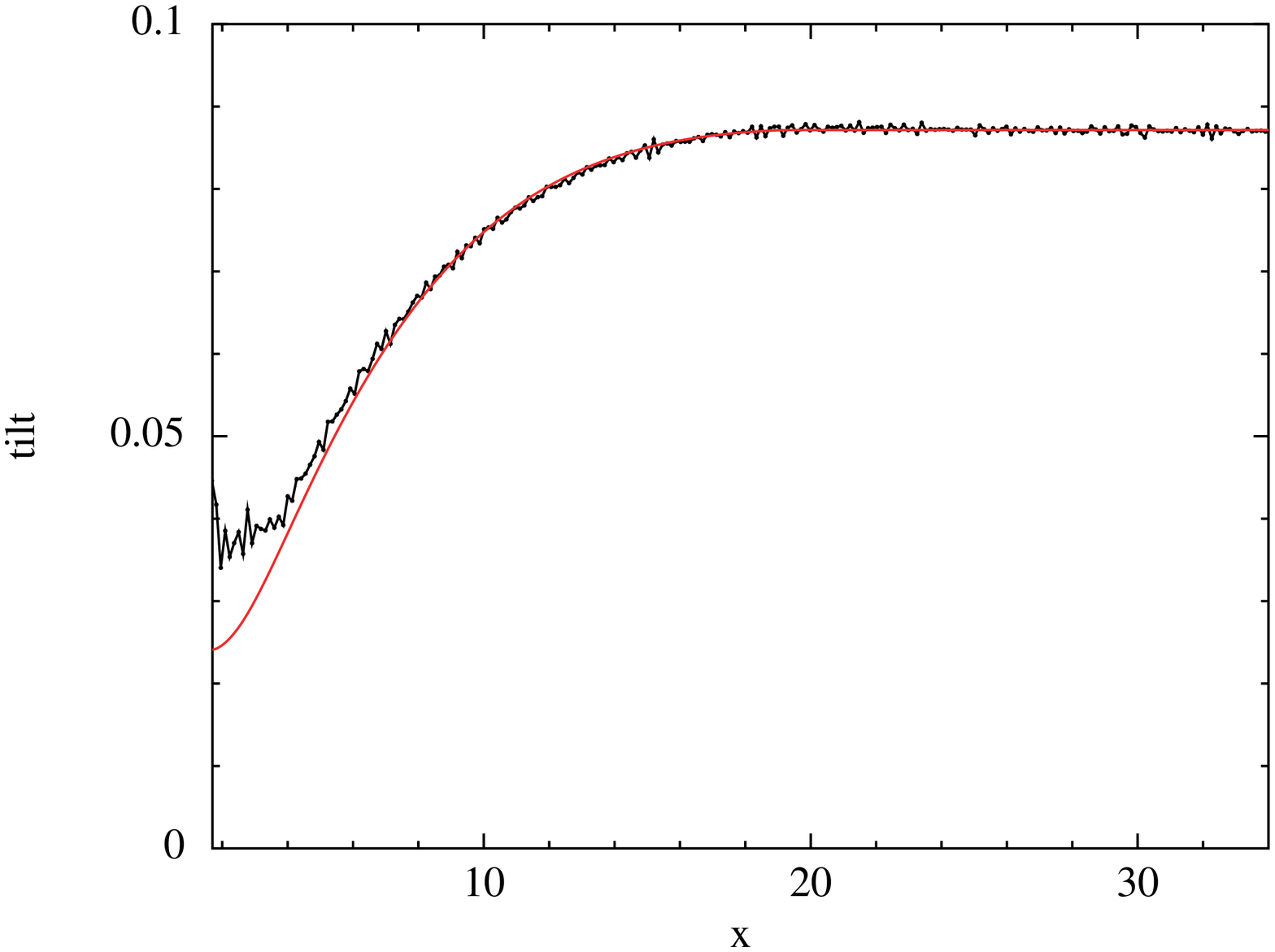}
\includegraphics[width=.9\columnwidth]{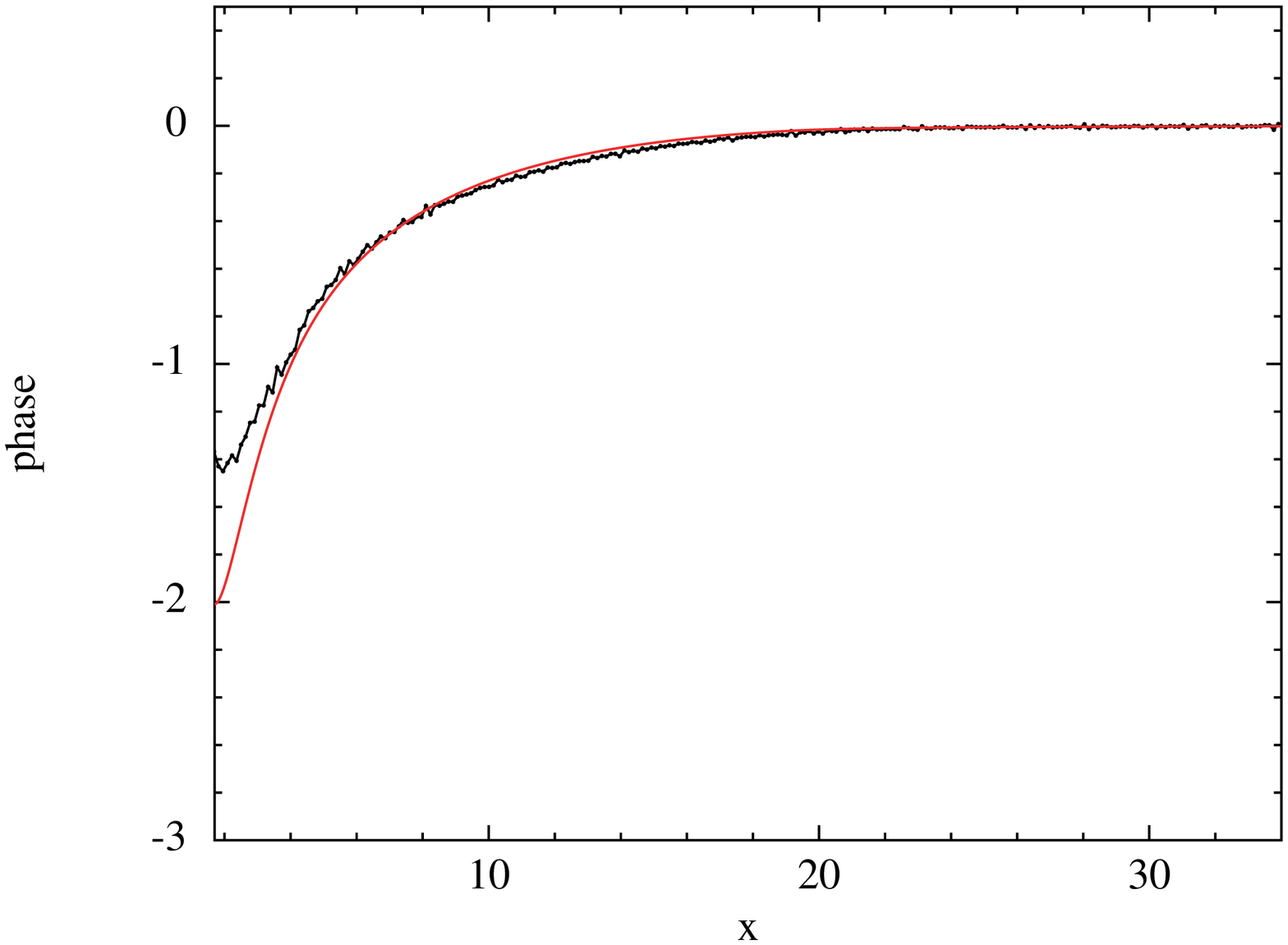}
\end{center}
\caption{Tilt and phase evolution of an initially untwisted tilted disc, with $\eta=0.25$, $(H_{\rm in}/R_{\rm in})=0.1$, $\alpha = 0.2$, $\beta_{\infty} = 5^\circ$ and $N = 1$ million at $t=2000$. Black lines show the SPH results, while red lines indicate the 1D model. The agreement is very good for viscous discs. Also here, the inner edge of the disc is set at $1.7a$ for the 1D runs. The figure is a nice confirmation that the equations do succeed in simulating the viscous regime on short timescales.}
\label{fig:res_02}
\end{figure*}

\begin{figure*}
\begin{center}
\includegraphics[width=.9\columnwidth]{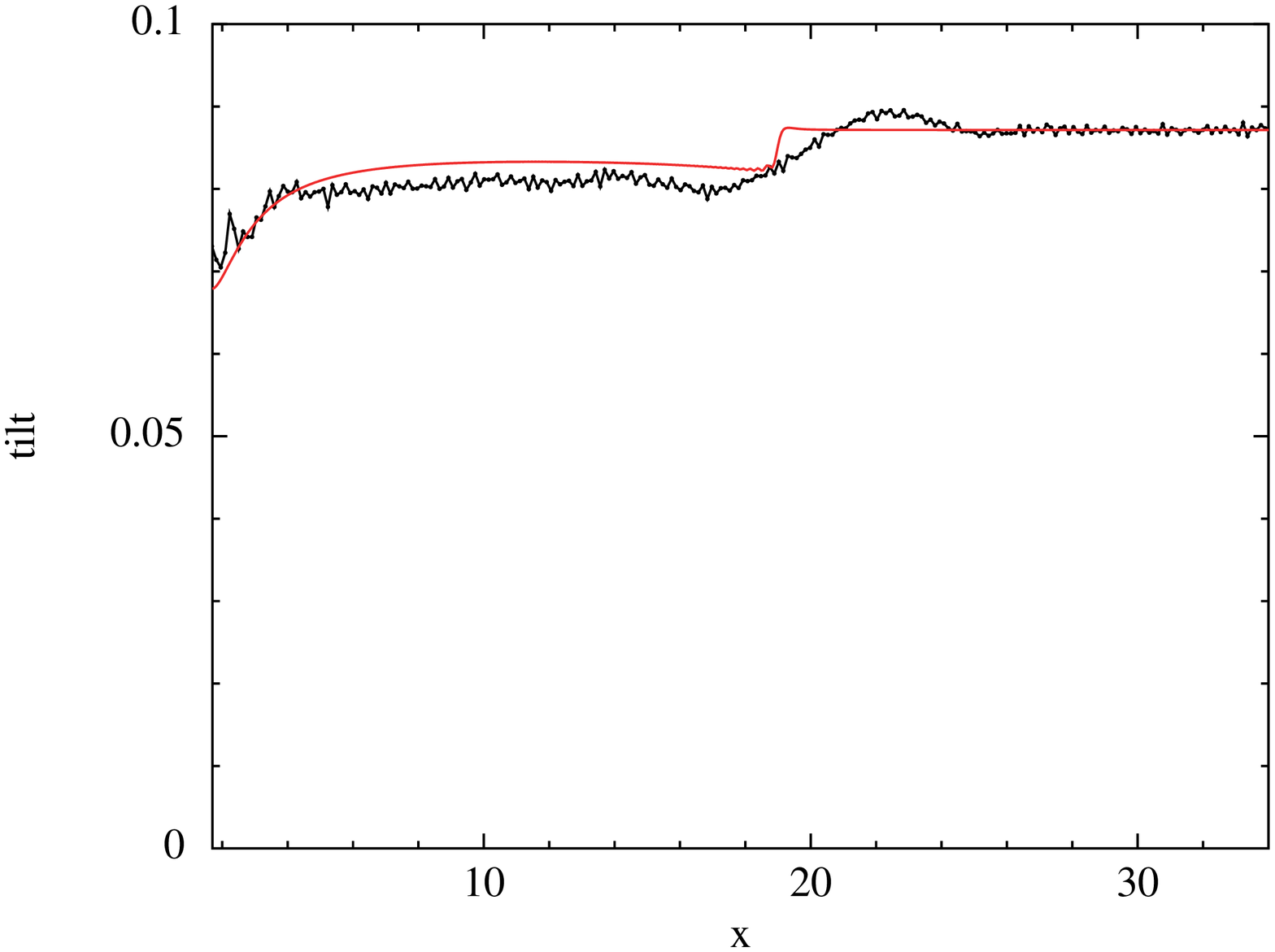}
\includegraphics[width=.9\columnwidth]{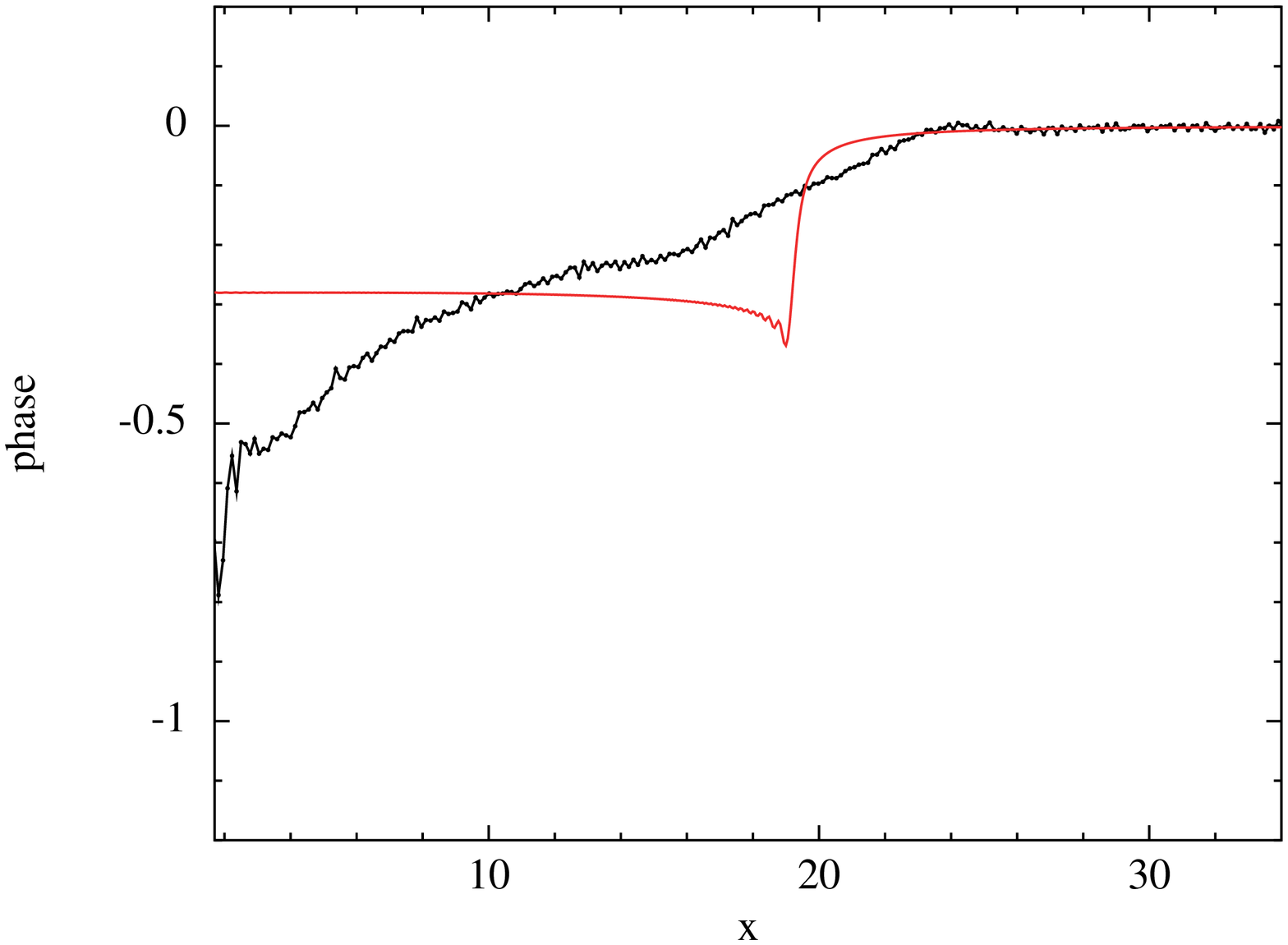}
\end{center}
\caption{Tilt and phase evolution of an initially untwisted tilted disc, with $\eta=0.25$, $(H_{\rm in}/R_{\rm in})=0.1$, $\alpha = 0$, $\beta_{\infty} = 5^\circ$ and $N = 1$ million at $t=2000$. Black lines show the SPH results, while red lines indicate the 1D model. Also here, the inner edge of the disc is set at $1.7a$ for the 1D runs. SPH succeeds in reproducing the tilt evolution, even for almost inviscid discs. The phase shows an apparent disagreement due to the contribution of the artificial viscosity to the effective one.}
\label{fig:res_00}
\end{figure*}

\subsubsection{Linear regime}

We analyse the warp dynamics in three cases: $\alpha = 0.05$, $\alpha = 0.2$ and $\alpha \approx 0$. In the first case the disc falls in the wave-like regime, in the second one it is in the diffusive regime, and the third one is the closest possible value to the inviscid case. Initially, at $t=0$, the disc is tilted with respect to the binary plane by an angle $\beta_{\infty} = 5^\circ$. Both \citet{nelson_pap99} and \citet{ogilvie06} have shown that in absence of an external torque (which does not modify the regime the waves propagate with, anyway) such a small inclination ensures a linear regime for the wave propagation. We have run other simulations with lower values of the initial $\beta_{\infty}$, but since we are dealing with full 3D simulations, we have to take the finite thickness into account. When $\beta_{\infty}$ is too small (i.e. $\tan{\beta_{\infty}} \lesssim H/R$), we obtain a very noisy measure of the tilt. By increasing $\beta_{\infty}$ up to $5^\circ$, we reduce the noise-to-signal ratio. Therefore we report the results of the $5^\circ$ simulations only.

Let us start with the case of $\alpha = 0.05$. In Fig. \ref{fig:res_005} we report the evolution of the tilt and the phase at $t = 2000$. The number of particles used in this simulation is $N=1$ million. We illustrate the results of the 3D simulation with the black lines, and the ones of the corresponding 1D simulation with the red lines. In the 1D simulations the inner edge of the disc has been set equal to $R_{\rm t}=1.7 a$. The general trend of the 3D simulations is in very good agreement with the 1D ones. Both the tilt and the twist tend to a steady state. Moreover, the tilt tends to a steady state that has a shape described by an evanescent wave, as discussed in section \ref{sec:waves_theory} theoretically, and in sections \ref{sec:analyt} and \ref{sec:1D} both analytically and numerically. There is a good agreement also in the propagation velocity. Overall, the evolution of both the tilt and the phase throughout the simulation is in very good agreement with that obtained by solving the linearised warp equations in 1D. 

An additional effect (not apparent in these plots) is that the evolution shows a small periodic oscillation with time. This is due to the imperfect rotational symmetry about the $z$-axis of the gravitational potential, which enforces wobbling modes into the disc \citep{bate00}, especially with $\sigma=2$.

We have also tested the effects of limited resolution in SPH by running a simulation with the same parameters as above but with 10 times fewer particles. The tilt evolution in this case is shown in Fig. \ref{fig:res_resolution}. We see that in this case the disc develops a smaller warp (i.e., the inner disc tends to stay more aligned with the outer disc). This is due to the fact that decreasing the resolution, we do not resolve equally well the inner disc edge. The disc thus appears truncated by tidal torques at a slightly larger radius ($\sim 2a$ in this case), thus decreasing the warp amplitude. We find a very good agreement in the evolution of the tilt by comparing this poorly resolved simulation with a 1D simulation with an inner radius equal to $2a$.

We have just described a disc in which the wave-like regime is expected in the whole disc. Now we use the same setup (simulations with $N = 1$ million particles) as above, but we set $\alpha = 0.2$. By knowing that $H/R = 0.1\ x^{-1/4}$, the condition $\alpha \gtrsim H/R$ is verified in the whole disc. Therefore with $\alpha = 0.2$ the warp evolves diffusively. We compare the 3D results with the ones obtained in section \ref{sec:1D} for the same value of $\alpha$. In Fig. \ref{fig:res_02} we report the results for the tilt and the phase evolution, respectively. The inner edge of the disc has been set to $R_{\rm t}=1.7 a$ for the 1D simulations. We observe a good agreement. This confirms that the equations do describe the evolution even in a diffusive regime (see section \ref{sec:waves_theory}). A small discrepancy is still present at the very inner edge because of the low resolution when $\Sigma$ tends to $0$ (and the associated error in the estimate of $R_{\rm t}$), but this does not affect the shape in the outer regions of the disc.

Finally, we try to simulate the dynamics of an inviscid disc. In order to do it, we consider a disc with physical viscosity equal to 0. However, as we have already mentioned in section \ref{sec:SPH}, we cannot remove the artificial viscosity from the simulations completely. In this section we use the following viscosity parameters: $\alpha_{\mathrm{art,max}} = 0.5$, $\alpha_{\mathrm{art,min}} = 10^{-5}$ and $\alpha = 0$. We used such a low value for $\alpha_{\mathrm{art,min}}$ because when ${\bf \nabla} \cdot {\bf v} < 0$ the Morris and Monaghan switch ensures that discontinuities are smoothed by a higher value of $\alpha_{\mathrm{art}}$. The artificial viscosity grows to its maximum value when \mbox{$|h{\bf \nabla} \cdot {\bf v}|>c_{\rm s}$}. We tried to use lower values of $\alpha_{\mathrm{art,max}}$, but, as predicted, the simulations become noisy. In Fig. \ref{fig:res_00} we report the tilt and the phase evolution of such a disc.

In the tilt we see a good agreement between the 3D simulation and the 1D solution (where, as usual, the inner radius of the disc has been set equal to $R_{\rm t}=1.7a$). In the 3D tilt we note a bump on the wavefront. This is due to the fact that in the 3D simulation $N$ is not large enough to resolve the tilt discontinuity at the wavefront of the 1D code: such a discontinuity is smoothed over some smoothing lengths.

The phase evolution is very different. As described in section \ref{sec:inviscid}, in the 1D case when $\alpha = 0$ the disc rotates and reaches a steady state that is untwisted. Instead, in the 3D case the disc does present a twist. This fact emphasises that our simulations are not completely inviscid, since the artificial viscosity is acting as a small effective viscosity in the disc (see equation \ref{eq:visc_sph}). The small amount of this effective viscosity produces the twist observable in the figure. Moreover, the resolution is not high enough to resolve the strong phase discontinuity of the 1D simulation. SPH is smoothing the discontinuity over some smoothing lengths.

Still, we note that the agreement between theory and simulations is remarkable, even for almost inviscid discs.

\begin{figure}
\centering
\includegraphics[width=.9\columnwidth]{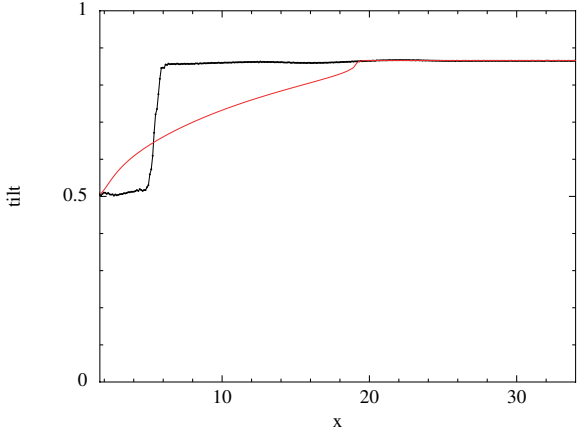}
\caption{Tilt and phase evolution of an initially untwisted tilted disc, with  $\beta_{\infty} = 60^\circ$, $\eta=0.25$, $(H_{\rm in}/R_{\rm in})=0.1$ and $N = 2$ million at $t=2000$. A sharp break occurs in the 3D simulations (black line). The linear theory (red line) fails to describe the warp evolution for such high inclinations.}
\label{fig:nonlinear_tilt}
\end{figure}

\begin{figure}
\centering
\includegraphics[width=.95\columnwidth]{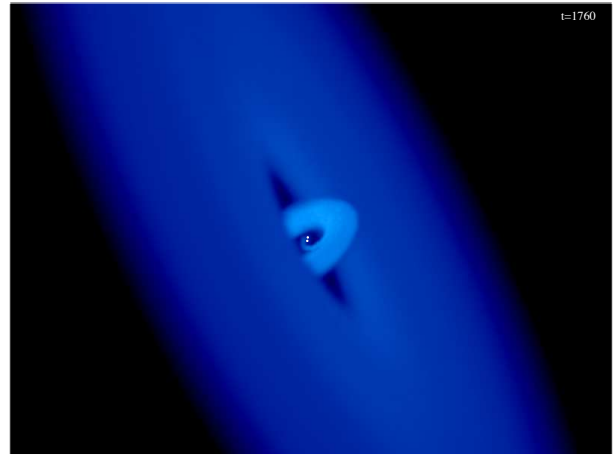}
\caption{3D structure of the disc shown in Fig. \ref{fig:nonlinear_tilt} at $t=1760$. The disc breaks in two almost separated discs. The inner one starts precessing, since its width is very narrow.}
\label{fig:nonlinear}
\end{figure}

\subsubsection{Non-linear regime}

If we enhance the initial inclination angle, the linear theory fails. As reported above, some efforts have been made in order to describe the non-linear regime, both numerically \citep{nelson_pap99} and analytically \citep{ogilvie06}, the latter in the absence of external torques. In this section we simulate the non-linear regime of wave-like warp propagation via SPH simulations with a much higher resolution than in previous works. Moreover, we focus on the case where the disc is subject to external torques, due to the central misaligned binary. We use $2$ million particles, and the following set of parameters: $\alpha_{\rm art,max}=0.5$, $\alpha_{\rm art,min}=0.01$, $\alpha=0.05$, $H_{\rm in}/R_{\rm in}=0.1$, $M_1=M_2=0.5$ and $r_1/R_{\rm in}=r_2/R_{\rm in}=0.5$. This is the same set used for the simulation portrayed in Fig. \ref{fig:res_005}.  We have used the usual formulation for the viscosity, since we know that in non-linear cases shocks are much more likely to occur.

We have performed simulations with three different initial inclinations of the disc plane with respect to the binary plane: $\beta_{\infty}=20$, $40$ and $60^\circ$. The $20^\circ$ inclined disc starts showing relevant discrepancies from the linear regime, but it is in the $40^\circ$ and even more in the $60^\circ$ inclined disc that the evolution is completely different from the one predicted by the linear theory. In this section we report the results of the most inclined disc only as an example. Further studies on this issue are required.

Let us analyse the case with $\beta_{\infty}=60^\circ$. In Fig. \ref{fig:nonlinear_tilt} we show the evolution of the tilt at $t=2000$. As above, we illustrate the results of the 3D simulation with the black line, and the ones corresponding to the linear 1D simulation with the red line. For the 1D simulation the inner edge has been set equal to $R_{\rm t}=1.7 a$ We can immediately observe that the discrepancy is very remarkable, and that the linear theory fails to describe the warp propagation in such an extreme case. From Fig. \ref{fig:nonlinear} we note that the disc breaks sharply. Moreover, the 3D simulation shows that the inner ring starts precessing.

This kind of behaviour has already been observed by \citet{lodato_price10} and \citet{nixon_al12}, the latter focusing on discs subject to Lense-Thirring precession. Their simulations focus on the diffusive regime ($\alpha>H/R$), but they both obtain the breaking of the disc with low values of $\alpha$. Moreover, \citet{fragner10} have seen the same result via a grid code, and \citet{larwood_pap97} via SPH simulations, but with a much lower resolution than ours. Note that we have performed the same kind of simulation, but with a lower viscosity ($\alpha=0.01$), and the general result is equivalent to the more viscous case.

\section{Conclusions}
\label{sec:concl}

In this paper we have analysed the bending-wave regime of protostellar circumbinary warped accretion discs. Analytically, we have found the general solution for the shape of an inclined disc around a binary of arbitrary mass ratio in the inviscid limit and in the linear approximation. We have verified that the solution for the steady state tilt of the disc is an evanescent wave, as predicted by \citetalias{LOP02} (where they obtained the same kind of solution for a retrograde rotating disc around a spinning black hole). In the inviscid limit, the disc does not present any twist, and the inner parts do not become aligned with the binary. This is different to the diffusive limit, where the disc tends to align in the inner regions.

Then, we have performed 1D time-dependent calculations for low viscosity discs affected by the binary torque. We can summarise our results as follows: firstly, we have verified that the warp does evolve as a bending wave with a wave front moving at $c_{\rm s}/2$. The disc reaches a steady state described by a stationary wave, and its shape does not depend on the initial condition. Internal torques due to a warped structure in the inner region compensate the differential precession that would occur if the rings forming the disc were disconnected. Secondly, we have explored the parameter space in terms of viscosity and amplitude of the external torque. We have found that the tilt shape of the steady state depends strongly on these two parameters. The amplitude of the external torque affects the amplitude of the warp in the inner regions. The smaller the mass ratio of the two stars (or the farther away from the disc they are), the smaller the amplitude of the warp. This fact is intuitive: as the mass ratio gets lower, the gravitational potential becomes more spherically symmetric. Therefore, the disc is less affected by an external torque. Instead, viscosity affects the disc in two ways. If the disc is viscous enough ($\alpha>H/R$), the warp propagates diffusively, as predicted by the theory, and aligns with the binary plane at the inner edge. Moreover, the amount of viscosity regulates the amplitude of the twist in the the disc. The more viscous the disc is, the more twisted it gets. Thirdly, we have run simulations with $\alpha=0$. In this last case, the disc reaches a steady state in very good agreement with the analytic solution. It tends to the same shape in the tilt, and to a constant phase that depends on the initial condition. However, we recall that the tilt normalisation and the constant phase angle are two degrees of freedom of the system, since we deal with a linear regime and rotational symmetry.

We can compare these results with the ones obtained by \citet{foucart13}, who also give approximate analytic estimates of the amplitude of the warp in the steady state shape of misaligned circumbinary discs. In our paper we have added significant contributions to their results. Firstly, we have explicitly derived an analytic solution for the inviscid case, in the form of modified Bessel functions. Secondly, we have analysed the temporal evolution of the shape of the discs, whereas \citet{foucart13} only focus on the steady state solutions, and we have explored the parameters space with time-dependent simulations. Thirdly, as they do, we show that the equations (and therefore the solutions) depend on two dimensionless parameters only, $\chi$ and $\alpha/(H_{\rm in}/R_{\rm in})$, and we broadly confirm their results on the amplitude of both the warping and the twisting. 

We have then performed 3D SPH simulations, in order to explore both the linear and the non-linear regime. We have compared these results with the ones obtained with the 1D ring code. By focusing on the linear case, we have firstly verified the validity of the assumptions made in the 1D model. Secondly, we have tested that the agreement is good, both in the tilt and in the twist, in the wave-like regime. Small discrepancies are due to the low resolution in the inner regions of the disc. By increasing the value of viscosity, we have verified that the wave equations do succeed in simulating the viscous regime (the agreement between 3D and 1D simulation is remarkable), with the \emph{caveat} that this is valid for short enough timescales. Finally, we have performed simulations at very low viscosities, close to the inviscid case. By comparing the SPH results to the 1D inviscid ones, we have obtained a good agreement, at least in the tilt evolution. This fact emphasises how well SPH is able to reproduce the warp evolution of discs, even for the case of extremely low viscosities. We have shown that standard tools, such as the Morris \& Monaghan switch, are indeed effective at reducing artificial viscosity and ensure the possibility of running almost inviscid warped disc SPH simulations. 

In the non-linear regime, we have shown that for high inclination angles of the disc plane with respect to the binary one the disc breaks, and the inner ring precesses almost completely disconnected from the outer regions of the disc. Additional studies are required to further explore this last issue.

\section*{Acknowledgements}
We thank Chris Nixon, Cathie Clarke and Jim Pringle for stimulating discussion. We thank the anonymous referee for useful advice and for pointing to us the paper by \citet{foucart13}. SF thanks the Science and Technology Facility Council and the Isaac Newton Trust for the award of a studentship. Figs \ref{fig:analytical}-\ref{fig:nonlinear} were produced using \textsc{splash} \citep{price07}, a visualisation tool for SPH data.

\appendix

\bibliography{warp_bib}

\begin{thebibliography}{}

\bibitem[\protect\citeauthoryear{{Albrecht}, {Winn}, {Johnson} \& {et
  al.}}{{Albrecht} et~al.}{2012}]{albrecht12}
{Albrecht} S.,  {Winn} J.~N.,  {Johnson} J.~A.,    {et al.} 2012, \apj, 757, 18

\bibitem[\protect\citeauthoryear{{Artymowicz} \& {Lubow}}{{Artymowicz} \&
  {Lubow}}{1994}]{art_lubow94}
{Artymowicz} P.,  {Lubow} S.~H.,  1994, \apj, 421, 651

\bibitem[\protect\citeauthoryear{{Bardeen} \& {Petterson}}{{Bardeen} \&
  {Petterson}}{1975}]{bardeen_petterson75}
{Bardeen} J.~M.,  {Petterson} J.~A.,  1975, \apjl, 195, L65

\bibitem[\protect\citeauthoryear{{Bate}, {Bonnell}, {Clarke} \& {et
  al.}}{{Bate} et~al.}{2000}]{bate00}
{Bate} M.~R.,  {Bonnell} I.~A.,  {Clarke} C.~J.,    {et al.} 2000, \mnras, 317,
  773

\bibitem[\protect\citeauthoryear{{Bate}, {Bonnell} \& {Price}}{{Bate}
  et~al.}{1995}]{bate95}
{Bate} M.~R.,  {Bonnell} I.~A.,    {Price} N.~M.,  1995, \mnras, 277, 362

\bibitem[\protect\citeauthoryear{{Bate}, {Lodato} \& {Pringle}}{{Bate}
  et~al.}{2010}]{bate10}
{Bate} M.~R.,  {Lodato} G.,    {Pringle} J.~E.,  2010, \mnras, 401, 1505

\bibitem[\protect\citeauthoryear{{Beuermann}, {Hessman}, {Dreizler} \& {et
  al.}}{{Beuermann} et~al.}{2010}]{beuermann10}
{Beuermann} K.,  {Hessman} F.~V.,  {Dreizler} S.,    {et al.} 2010, \aap, 521,
  L60

\bibitem[\protect\citeauthoryear{{Beust} \& {Dutrey}}{{Beust} \&
  {Dutrey}}{2005}]{beust_dutrey05}
{Beust} H.,  {Dutrey} A.,  2005, \aap, 439, 585

\bibitem[\protect\citeauthoryear{{Chiang} \& {Murray-Clay}}{{Chiang} \&
  {Murray-Clay}}{2004}]{CM04}
{Chiang} E.~I.,  {Murray-Clay} R.~A.,  2004, \apj, 607, 913

\bibitem[\protect\citeauthoryear{{Deeg}, {Oca{\~n}a}, {Kozhevnikov} \& {et
  al.}}{{Deeg} et~al.}{2008}]{deeg08}
{Deeg} H.~J.,  {Oca{\~n}a} B.,  {Kozhevnikov} V.~P.,    {et al.} 2008, \aap,
  480, 563

\bibitem[\protect\citeauthoryear{{Demianski} \& {Ivanov}}{{Demianski} \&
  {Ivanov}}{1997}]{demianski_ivanov97}
{Demianski} M.,  {Ivanov} P.~B.,  1997, \aap, 324, 829

\bibitem[\protect\citeauthoryear{{Dotti}, {Volonteri}, {Perego} \& {et
  al.}}{{Dotti} et~al.}{2010}]{dotti10}
{Dotti} M.,  {Volonteri} M.,  {Perego} A.,    {et al.} 2010, \mnras, 402, 682

\bibitem[\protect\citeauthoryear{{Doyle}, {Carter}, {Fabrycky} \& {et
  al.}}{{Doyle} et~al.}{2011}]{doyle11}
{Doyle} L.~R.,  {Carter} J.~A.,  {Fabrycky} D.~C.,    {et al.} 2011, Science,
  333, 1602

\bibitem[\protect\citeauthoryear{{Dutrey}, {Guilloteau} \& {Simon}}{{Dutrey}
  et~al.}{1994}]{dutrey94}
{Dutrey} A.,  {Guilloteau} S.,    {Simon} M.,  1994, \aap, 286, 149

\bibitem[\protect\citeauthoryear{{Espa{\~n}ol} \& {Revenga}}{{Espa{\~n}ol} \&
  {Revenga}}{2003}]{espanol_revenga07}
{Espa{\~n}ol} P.,  {Revenga} M.,  2003, \pre, 67, 026705

\bibitem[\protect\citeauthoryear{{Flebbe}, {Muenzel}, {Herold} \& {et
  al.}}{{Flebbe} et~al.}{1994}]{flebbe94}
{Flebbe} O.,  {Muenzel} S.,  {Herold} H.,    {et al.} 1994, \apj, 431, 754

\bibitem[\protect\citeauthoryear{{Foucart} \& {Lai}}{{Foucart} \&
  {Lai}}{2013}]{foucart13}
{Foucart} F.,  {Lai} D.,  2013, \apj, 764, 106

\bibitem[\protect\citeauthoryear{{Fragner} \& {Nelson}}{{Fragner} \&
  {Nelson}}{2010}]{fragner10}
{Fragner} M.~M.,  {Nelson} R.~P.,  2010, \aap, 511, A77

\bibitem[\protect\citeauthoryear{{Gammie}, {Goodman} \& {Ogilvie}}{{Gammie}
  et~al.}{2000}]{gammie00}
{Gammie} C.~F.,  {Goodman} J.,    {Ogilvie} G.~I.,  2000, \mnras, 318, 1005

\bibitem[\protect\citeauthoryear{{Herrnstein}, {Greenhill} \&
  {Moran}}{{Herrnstein} et~al.}{1996}]{herrnstein96}
{Herrnstein} J.~R.,  {Greenhill} L.~J.,    {Moran} J.~M.,  1996, \apjl, 468,
  L17

\bibitem[\protect\citeauthoryear{{Hjellming} \& {Rupen}}{{Hjellming} \&
  {Rupen}}{1995}]{hjellming95}
{Hjellming} R.~M.,  {Rupen} M.~P.,  1995, \nat, 375, 464

\bibitem[\protect\citeauthoryear{{Ivanov}, {Papaloizou} \& {Polnarev}}{{Ivanov}
  et~al.}{1999}]{ivanov99}
{Ivanov} P.~B.,  {Papaloizou} J.~C.~B.,    {Polnarev} A.~G.,  1999, \mnras,
  307, 79

\bibitem[\protect\citeauthoryear{{King}, {Pringle} \& {Hofmann}}{{King}
  et~al.}{2008}]{king08}
{King} A.~R.,  {Pringle} J.~E.,    {Hofmann} J.~A.,  2008, \mnras, 385, 1621

\bibitem[\protect\citeauthoryear{{Larwood} \& {Papaloizou}}{{Larwood} \&
  {Papaloizou}}{1997}]{larwood_pap97}
{Larwood} J.~D.,  {Papaloizou} J.~C.~B.,  1997, \mnras, 285, 288

\bibitem[\protect\citeauthoryear{{Lee}, {Kim}, {Kim} \& {et al.}}{{Lee}
  et~al.}{2009}]{lee09}
{Lee} J.~W.,  {Kim} S.-L.,  {Kim} C.-H.,    {et al.} 2009, \aj, 137, 3181

\bibitem[\protect\citeauthoryear{{Lodato} \& {Facchini}}{{Lodato} \&
  {Facchini}}{2013}]{facchini13_2}
{Lodato} G.,  {Facchini} S.,  2013, \mnras, accepted for publication

\bibitem[\protect\citeauthoryear{{Lodato} \& {Gerosa}}{{Lodato} \&
  {Gerosa}}{2012}]{lodato_gerosa12}
{Lodato} G.,  {Gerosa} D.,  2012, \mnras, p.~L14

\bibitem[\protect\citeauthoryear{{Lodato} \& {Price}}{{Lodato} \&
  {Price}}{2010}]{lodato_price10}
{Lodato} G.,  {Price} D.~J.,  2010, \mnras, 405, 1212

\bibitem[\protect\citeauthoryear{{Lodato} \& {Pringle}}{{Lodato} \&
  {Pringle}}{2006}]{lodato_pringle06}
{Lodato} G.,  {Pringle} J.~E.,  2006, \mnras, 368, 1196

\bibitem[\protect\citeauthoryear{{Lodato} \& {Pringle}}{{Lodato} \&
  {Pringle}}{2007}]{lodato_pringle07}
{Lodato} G.,  {Pringle} J.~E.,  2007, \mnras, 381, 1287

\bibitem[\protect\citeauthoryear{{Lubow} \& {Ogilvie}}{{Lubow} \&
  {Ogilvie}}{2000}]{lubow_ogilvie2000}
{Lubow} S.~H.,  {Ogilvie} G.~I.,  2000, \apj, 538, 326

\bibitem[\protect\citeauthoryear{{Lubow} \& {Ogilvie}}{{Lubow} \&
  {Ogilvie}}{2001}]{lubow_ogilvie01}
{Lubow} S.~H.,  {Ogilvie} G.~I.,  2001, \apj, 560, 997

\bibitem[\protect\citeauthoryear{{Lubow}, {Ogilvie} \& {Pringle}}{{Lubow}
  et~al.}{2002}]{LOP02}
{Lubow} S.~H.,  {Ogilvie} G.~I.,    {Pringle} J.~E.,  2002, \mnras, 337, 706

\bibitem[\protect\citeauthoryear{{Martin}, {Tout} \& {Pringle}}{{Martin}
  et~al.}{2008}]{martin08}
{Martin} R.~G.,  {Tout} C.~A.,    {Pringle} J.~E.,  2008, \mnras, 387, 188

\bibitem[\protect\citeauthoryear{{Mathieu}, {Stassun}, {Basri} \& {et
  al.}}{{Mathieu} et~al.}{1997}]{mathieu97}
{Mathieu} R.~D.,  {Stassun} K.,  {Basri} G.,    {et al.} 1997, \aj, 113, 1841

\bibitem[\protect\citeauthoryear{{McKee} \& {Ostriker}}{{McKee} \&
  {Ostriker}}{2007}]{mckee_ostriker07}
{McKee} C.~F.,  {Ostriker} E.~C.,  2007, \araa, 45, 565

\bibitem[\protect\citeauthoryear{{Morris} \& {Monaghan}}{{Morris} \&
  {Monaghan}}{1997}]{morris_mon97}
{Morris} J.~P.,  {Monaghan} J.~J.,  1997, J. Comp. Phys., 136, 41

\bibitem[\protect\citeauthoryear{{Murray}}{{Murray}}{1996}]{murray96}
{Murray} J.~R.,  1996, \mnras, 279, 402

\bibitem[\protect\citeauthoryear{{Nayakshin}}{{Nayakshin}}{2005}]{naya05}
{Nayakshin} S.,  2005, \mnras, 359, 545

\bibitem[\protect\citeauthoryear{{Nelson} \& {Papaloizou}}{{Nelson} \&
  {Papaloizou}}{1999}]{nelson_pap99}
{Nelson} R.~P.,  {Papaloizou} J.~C.~B.,  1999, \mnras, 309, 929

\bibitem[\protect\citeauthoryear{{Nelson} \& {Papaloizou}}{{Nelson} \&
  {Papaloizou}}{2000}]{nelson_pap00}
{Nelson} R.~P.,  {Papaloizou} J.~C.~B.,  2000, \mnras, 315, 570

\bibitem[\protect\citeauthoryear{{Nixon}, {King}, {Price} \& {Frank}}{{Nixon}
  et~al.}{2012}]{nixon_al12}
{Nixon} C.,  {King} A.,  {Price} D.,    {Frank} J.,  2012, \apjl, 757, L24

\bibitem[\protect\citeauthoryear{{Nixon}}{{Nixon}}{2012}]{nixon2012}
{Nixon} C.~J.,  2012, \mnras, 423, 2597

\bibitem[\protect\citeauthoryear{{Nixon} \& {King}}{{Nixon} \&
  {King}}{2012}]{nixon_king12}
{Nixon} C.~J.,  {King} A.~R.,  2012, \mnras, 421, 1201

\bibitem[\protect\citeauthoryear{{Nixon}, {King} \& {Pringle}}{{Nixon}
  et~al.}{2011}]{nixon2011}
{Nixon} C.~J.,  {King} A.~R.,    {Pringle} J.~E.,  2011, \mnras, 417, L66

\bibitem[\protect\citeauthoryear{{Ogilvie}}{{Ogilvie}}{1999}]{ogilvie99}
{Ogilvie} G.~I.,  1999, \mnras, 304, 557

\bibitem[\protect\citeauthoryear{{Ogilvie}}{{Ogilvie}}{2006}]{ogilvie06}
{Ogilvie} G.~I.,  2006, \mnras, 365, 977

\bibitem[\protect\citeauthoryear{{Orosz}, {Welsh}, {Carter} \& {et
  al.}}{{Orosz} et~al.}{2012}]{orosz12}
{Orosz} J.~A.,  {Welsh} W.~F.,  {Carter} J.~A.,    {et al.} 2012, Science, 337,
  1511

\bibitem[\protect\citeauthoryear{{Papaloizou} \& {Lin}}{{Papaloizou} \&
  {Lin}}{1995}]{papaloizou_lin95}
{Papaloizou} J.~C.~B.,  {Lin} D.~N.~C.,  1995, \apj, 438, 841

\bibitem[\protect\citeauthoryear{{Papaloizou} \& {Pringle}}{{Papaloizou} \&
  {Pringle}}{1983}]{papaloizou_pringle83}
{Papaloizou} J.~C.~B.,  {Pringle} J.~E.,  1983, \mnras, 202, 1181

\bibitem[\protect\citeauthoryear{{Price}}{{Price}}{2007}]{price07}
{Price} D.~J.,  2007, \pasa, 24, 159

\bibitem[\protect\citeauthoryear{{Price}}{{Price}}{2012}]{price12}
{Price} D.~J.,  2012, J. Comp. Phys., 231, 759

\bibitem[\protect\citeauthoryear{{Price} \& {Federrath}}{{Price} \&
  {Federrath}}{2010}]{price_fed10}
{Price} D.~J.,  {Federrath} C.,  2010, \mnras, 406, 1659

\bibitem[\protect\citeauthoryear{{Pringle}}{{Pringle}}{1992}]{pringle92}
{Pringle} J.~E.,  1992, \mnras, 258, 811

\bibitem[\protect\citeauthoryear{{Pringle}}{{Pringle}}{1996}]{pringle96}
{Pringle} J.~E.,  1996, \mnras, 281, 357

\bibitem[\protect\citeauthoryear{{Scheuer} \& {Feiler}}{{Scheuer} \&
  {Feiler}}{1996}]{scheuer_feiler96}
{Scheuer} P.~A.~G.,  {Feiler} R.,  1996, \mnras, 282, 291

\bibitem[\protect\citeauthoryear{{Shakura} \& {Sunyaev}}{{Shakura} \&
  {Sunyaev}}{1973}]{shakura73}
{Shakura} N.~I.,  {Sunyaev} R.~A.,  1973, \aap, 24, 337

\bibitem[\protect\citeauthoryear{{Tananbaum}, {Gursky}, {Kellogg}, {Levinson},
  {Schreier} \& {Giacconi}}{{Tananbaum} et~al.}{1972}]{tananbaum72}
{Tananbaum} H.,  {Gursky} H.,  {Kellogg} E.~M.,  {Levinson} R.,  {Schreier} E.,
     {Giacconi} R.,  1972, \apjl, 174, L143

\bibitem[\protect\citeauthoryear{{Triaud}, {Collier Cameron}, {Queloz} \& {et
  al.}}{{Triaud} et~al.}{2010}]{triaud10}
{Triaud} A.~H.~M.~J.,  {Collier Cameron} A.,  {Queloz} D.,    {et al.} 2010,
  \aap, 524, A25

\bibitem[\protect\citeauthoryear{{von Neumann} \& {Richtmyer}}{{von Neumann} \&
  {Richtmyer}}{1950}]{von50}
{von Neumann} J.,  {Richtmyer} R.~D.,  1950, J. App. Phys., 21, 232

\bibitem[\protect\citeauthoryear{{Welsh}, {Orosz}, {Carter} \& {et
  al.}}{{Welsh} et~al.}{2012}]{welsh12}
{Welsh} W.~F.,  {Orosz} J.~A.,  {Carter} J.~A.,    {et al.} 2012, \nat, 481,
  475

\bibitem[\protect\citeauthoryear{{Wijers} \& {Pringle}}{{Wijers} \&
  {Pringle}}{1999}]{wijers_pringle99}
{Wijers} R.~A.~M.~J.,  {Pringle} J.~E.,  1999, \mnras, 308, 207

\end{thebibliography}

\label{lastpage}
\end{document}